\documentclass[twocolumn ]{aastex631}

\shorttitle{AASTeX v6.3.1 Sample article}
\shortauthors{Ru\'iz-Rodr\'iguez et al.}
\graphicspath{{./}{figures/}}

\begin{document}

\title{Discovery of a brown dwarf with quasi-spherical mass-loss}

\author[0000-0003-3573-8163]{Dary A. Ru\'iz-Rodr\'iguez}\footnote{Dary Ru\'iz-Rodr\'iguez is a Jansky Fellow of the National Radio Astronomy Observatory.}
\affiliation{National Radio Astronomy Observatory; 520 Edgemont Rd.,
 Charlottesville, VA 22903, USA}

\author[0000-0002-2828-1153]{Lucas A. Cieza}
\affiliation{Facultad de Ingenier\'ia y Ciencias, N\'ucleo de Astronom\'ia, Universidad Diego Portales, Av. Ejercito 441. Santiago, Chile}

\affiliation{Millennium Nucleus on Young Exoplanets and their Moons (YEMS), Universidad Diego Portales, Av. Ejercito 441, Santiago, Chile}



\author[0000-0002-0433-9840]{Simon Casassus}
\affiliation{Departamento de Astronom\'{\i}a, Universidad de Chile, Casilla 36-D, Santiago, Chile}
\affiliation{Facultad de Ingenier\'ia y Ciencias, Universidad Adolfo Ib\'a\~nez, Av. Diagonal las Torres 2640, Pe\~{n}alol\'{e}n, Santiago, Chile }
\affiliation{Data Observatory Foundation, Universidad de Chile, Casilla 36-D, Santiago, Chile}
\affiliation{Millennium Nucleus on Young Exoplanets and their Moons  (YEMS), Universidad de Chile, Casilla 36-D, Santiago, Chile}

\author{Victor Almendros-Abad}
\affiliation{CENTRA/SIM, Faculdade de Ciencias de Universidade de Lisboa;  Ed. C8, Campo Grande, P-1749-016 Lisboa, Portugal}

\author{Paula Jofr\'e}
\affiliation{Facultad de Ingenier\'ia y Ciencias, N\'ucleo de Astronom\'ia, Universidad Diego Portales, Av. Ejercito 441. Santiago, Chile}

\author{Koraljka Muzic}
\affiliation{CENTRA/SIM, Faculdade de Ciencias de Universidade de Lisboa;  Ed. C8, Campo Grande, P-1749-016 Lisboa, Portugal}

\author{Karla Pe\~na Ramirez}
\affiliation{Centro de Astronom\'ia (CITEVA), Universidad de Antofagasta; Av. Angamos 601, Antofagasta 1240000, Chile}

\author{Grace Batalla-Falcon}
\affiliation{Facultad de Ingenier\'ia y Ciencias, N\'ucleo de Astronom\'ia, Universidad Diego Portales, Av. Ejercito 441. Santiago, Chile}

\author{Michael M. Dunham}
\affiliation{Department of Physics, State University of New York Fredonia; Fredonia, NY 14063, USA}

\author{Camilo Gonz\'alez-Ruilova}
\affiliation{Facultad de Ingenier\'ia y Ciencias, N\'ucleo de Astronom\'ia, Universidad Diego Portales, Av. Ejercito 441. Santiago, Chile}
\affiliation{European Southern Observatory; Alonso de Cordova 3107, Vitacura, Casilla, 19001, Santiago, Chile}

\author{Antonio Hales}
\affiliation{Joint ALMA Observatory; Alonso de Cordova 3107, Vitacura 763-0355, Santiago, Chile}

\author{Elizabeth Humphreys}
\affiliation{Joint ALMA Observatory; Alonso de Cordova 3107, Vitacura 763-0355, Santiago, Chile}

\author{Pedro H., Nogueira}
\affiliation{Facultad de Ingenier\'ia y Ciencias, N\'ucleo de Astronom\'ia, Universidad Diego Portales, Av. Ejercito 441. Santiago, Chile}

\author{Claudia Paladini}
\affiliation{European Southern Observatory; Alonso de Cordova 3107, Vitacura, Casilla, 19001, Santiago, Chile}

\author{John Tobin}
\affiliation{National Radio Astronomy Observatory; 520 Edgemont Rd.,
 Charlottesville, VA 22903, USA}

 \author{Jonathan P. Williams}
\affiliation{Institute for Astronomy, University of Hawaii at Manoa; Woodlawn Drive, Honolulu, HI, 96822, USA}

\author{Alice Zurlo}
\affiliation{Facultad de Ingenier\'ia y Ciencias, N\'ucleo de Astronom\'ia, Universidad Diego Portales, Av. Ejercito 441. Santiago, Chile}

\affiliation{Millennium Nucleus on Young Exoplanets and their Moons (YEMS), Universidad Diego Portales, Av. Ejercito 441, Santiago, Chile}



\begin{abstract}

We report the serendipitous discovery of an elliptical shell of CO associated with the faint stellar object SSTc2d J163134.1-24006 as part of the ``Ophiuchus Disk Survey Employing ALMA" (ODISEA), a project aiming to study the entire population of protoplanetary disks in the Ophiuchus Molecular Cloud from 230 GHz continuum emission and  $^{12}$CO (J=2-1), $^{13}$CO (J=2-1) and C$^{18}$CO (J=2-1) lines readable in Band-6. Remarkably, we detect a bright $^{12}$CO elliptical shape emission of $\sim$ 3$^{"}$ $\times$ 4$^{"}$ towards SSTc2d J163134.1-24006 without a 230 GHz continuum detection. Based on the observed near-IR spectrum taken with the Very Large Telescope (KMOS), the brightness of the source,  its 3-dimensional motion, and Galactic dynamic arguments, we conclude that the source is not a giant star in the distant background ($>$5 - 10 kpc) and is most likely to be a young brown dwarf in the Ophiuchus cloud, at a distance of just $\sim$139 pc. This is the first report of  quasi-spherical mass loss in a young brown dwarf. We suggest that the observed shell could be associated with a thermal pulse produced by the fusion of deuterium, which is not yet well understood,  but for a sub-stellar object is expected to occur during a short period of time at an age of a few Myr, in agreement with the ages of the objects in the region. Other  more exotic scenarios, such as a merger with planetary companions, cannot be ruled out from the current observations.

\end{abstract}



\section{Introduction} \label{Sec:Intro}

The internal structure and evolution of very low-mass stars (VLMS) and brown dwarfs (BDs) are not well understood yet. According to the definition of BDs, these objects can burn deuterium above the limiting mass of $\sim$13 $\rm M_{Jup}$ \citep{Burrows2001, Chabrier2005}, but are unable to burn regular hydrogen, which requires a minimum mass of 80 M$\rm _{Jup}$ and a core temperature of $\sim$ 3$\times$10$^{6}$ K. The deuterium fusion mass depends on different factors such as the initial deuterium abundance, and object's metallicity \citep{Chabrier1997, Spiegel2011}. In the case of BDs with $\lesssim$ 0.1 $\rm M_{\odot}$, their internal structures are expected to be fully convective while undergoing a state of quasi-static contraction in order to start burning deuterium. This contraction process is slowed or even halted until the supply of deuterium is depleted. Such approximation is only justified as the core temperature constantly rises during the contraction at the end of the accretion process. As a consequence, BDs could become pulsationally unstable during the burning phase \citep{Palla2005}. In this context, wind variations associated with thermal pulses and sudden increases of accretion rates could have indirect effects on the evolution of the system. However, there has not been observational evidence or more detailed theoretical work related to the effects of such ``thermal runaway pulsation" and/or sudden increases in the accretion rates (e.g. planet consumption by the host star) in low mass stars at the earliest formation phases.

Mass-loss in the form of expanding shells of gas is a well-known characteristic feature of the late evolution of stars with masses ranging from ~1 to 4 $\rm M_{\odot}$ \citep[e.g.;][]{Vassiliadis1993}, after they have consumed the hydrogen at their cores and leave the main sequence to become giants \citep{Iben1983}. Some of the largest mass-loss events are associated with thermal pulses originating from the rapid fusion of shells of Helium in the interior of stars in the asymptotic giant branch \citep{ Maercker2012}. Thus far, an analogous phenomenon has not been observed in pre-main-sequence stars or young brown dwarfs (mass $<$ 0.08 $\rm M_{\odot}$, age $<$ 100 Myr).

Here we present the serendipitous discovery of an external CO shell associated with a faint stellar object (SSTc2d J163134.1-24006) while performing the ``Ophiuchus Disk Survey Employing ALMA" \citep[ODISEA;][]{Cieza2019}. Given the very puzzling nature of the system (a faint IR source with strong CO emission lacking a mm detection), we have considered several scenarios, including a first hydro-static core, a red giant in the background and young brown dwarf in Ophiuchus. From current observational constraints, we have concluded that the observed CO elliptical-like shape emission (``bubble") with ALMA may be the result of a a very-low-mass object with spherical mass loss. The structure of this article is as follows: Properties of SSTc2d J163134.1-24006 are introduced in Sec \ref{Sec:Target}. In Sec. \ref{Sec:Data}, we present IR and milimeter observations taken with the VLT and ALMA towards SSTc2d J163134.1-24006. The analysis carried out and main results from these observations are described in Sec. \ref{Sec:Analysis}. Section \ref{Sec:Scenario} describes the explored astrophysical phenomena that might explain the mystery object together with pros and cons, while Sec. \ref{Sec:Model} presents the most likely scenario that describes our discovery. Finally, we discuss the proposed stellar scenario in Sec. \ref{Sec:Discussion}. A summary of our work can be found in Sec.\ref{Sec:Summary}.

\section{Target} \label{Sec:Target}

\begin{figure*}[ht!]
\epsscale{0.65}
\plotone{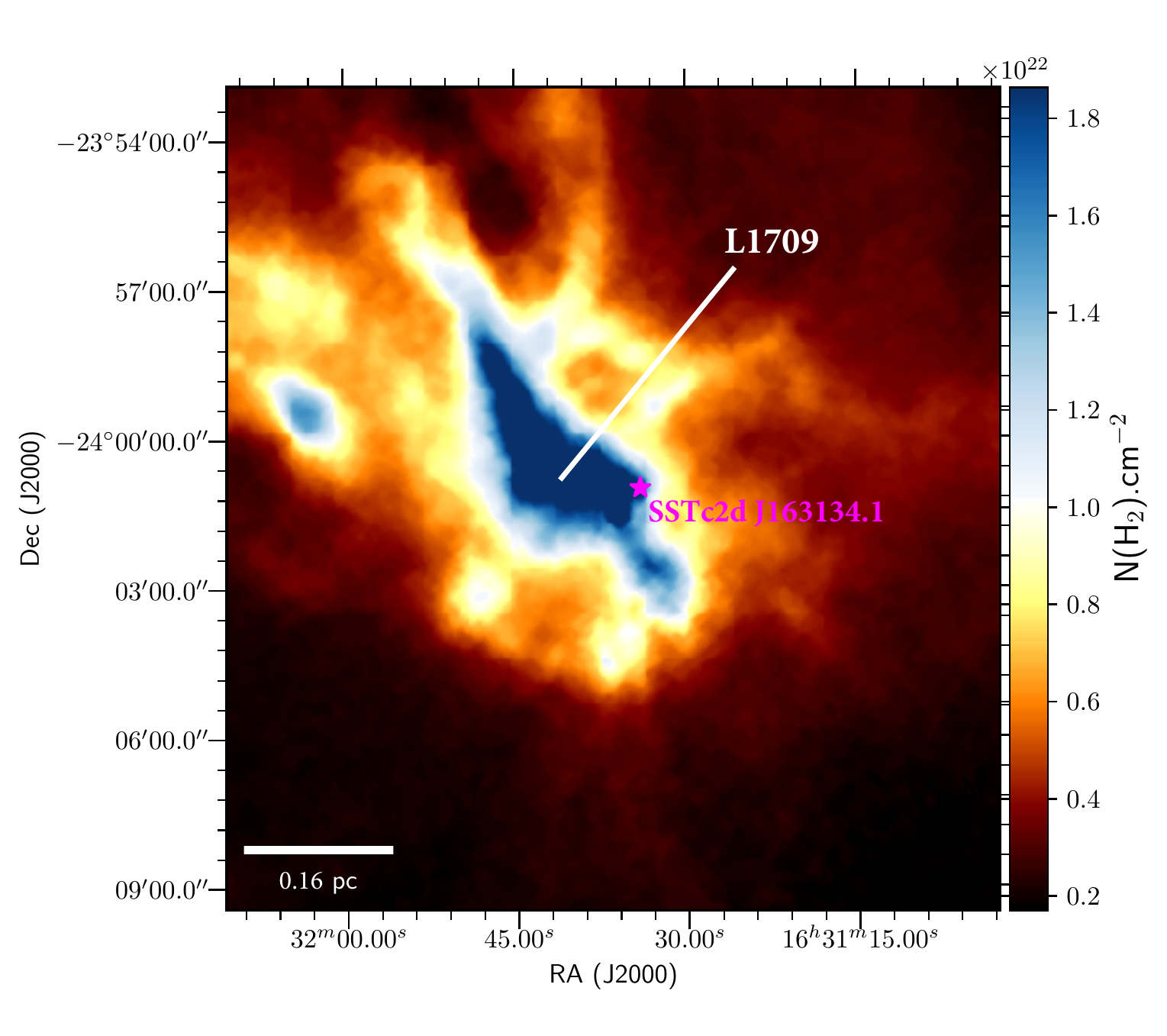}
\caption{Herschel column-density map of the Ophiuchus molecular cloud at an effective resolution of 18.2$^{"}$. The column-density map is given in units of H$_{2}$ cm$^{-2}$ derived by \citet{Ladjelate2020}. The magenta star indicates the location of the near IR source SSTc2d J163134.1 in a region with a relatively high H$_{2}$ column density of $\sim$2.5$\times$10$^{22}$ cm $^{-2}$. The Lynds L1709 dark cloud in the region is indicated. The bar at the lower right indicates a scale of $\sim$0.16 pc assuming a distance of 140 pc to the entire region \citep{Loinard2008}.}
\label{Fig:Density}
\end{figure*}

The quasi-sphere of gas discovered is associated with the candidate Young Stellar Object (YSO) SSTc2d J163134.1-240060 (SSTc2d J163134.1 hereafter). Based on its spatial location and projection, SSTc2d J163134.1 likely is a member of the Ophiuchus complex dark cloud L1709 (Fig. \ref{Fig:Density}), suggesting a median distance of $\sim$139 pc and an age of $\sim$ 2 Myr \citep{Esplin2020}. SSTc2d J163134.1 was initially identified by the Spitzer Legacy Project “Cores to Disks” based on its infrared (IR) Spectral Energy Distribution (SED) \citep{Evans2009}. In the infrared, and further discussed below, SSTc2d J163134.1 is detected between 2.2 and 24 $\micron$ with an SED peaking at 4.5 $\micron$. The spectral slope ($\rm \alpha_{IR}$) of SSTc2d J163134.1 calculated by \citet{Evans2009} between 2.2 and 24 $\micron$ is -0.32. This $\rm \alpha_{IR}$ value indicates a Class II source in the standard YSO classification \citep{Williams2011}, which in principle corresponds to an YSO surrounded by a disk but without a significant molecular envelope. However, the observed slope can be biased to higher values in regions of strong extinction. Interestingly, the Herschel column density map (Fig. \ref{Fig:Density}) at 18.2$^{''}$ resolution indicates an H$_{2}$ column density of 2.5$\times$10$^{22}$ cm $^{-2}$ at the location of the source, corresponding to Av $\sim$ 36 mag along the line of sight \citep{Ladjelate2020}. The reddening produced by such a high extinction could increase the IR value, mimicking the presence of mid-IR excess. If the large IR value is due to high reddening instead of the thermal emission from a disk, this would imply that  SSTc2d J163134.1 was observed with ALMA by chance as the ODISEA program aimed to observe YSOs surrounded by a disk \citep{Cieza2019, Williams2019}.

\section{Observations and data reduction} \label{Sec:Data}

\subsection{ALMA}
\label{Sec:ALMA}

\begin{figure*}[ht!]
\epsscale{0.85}
\plotone{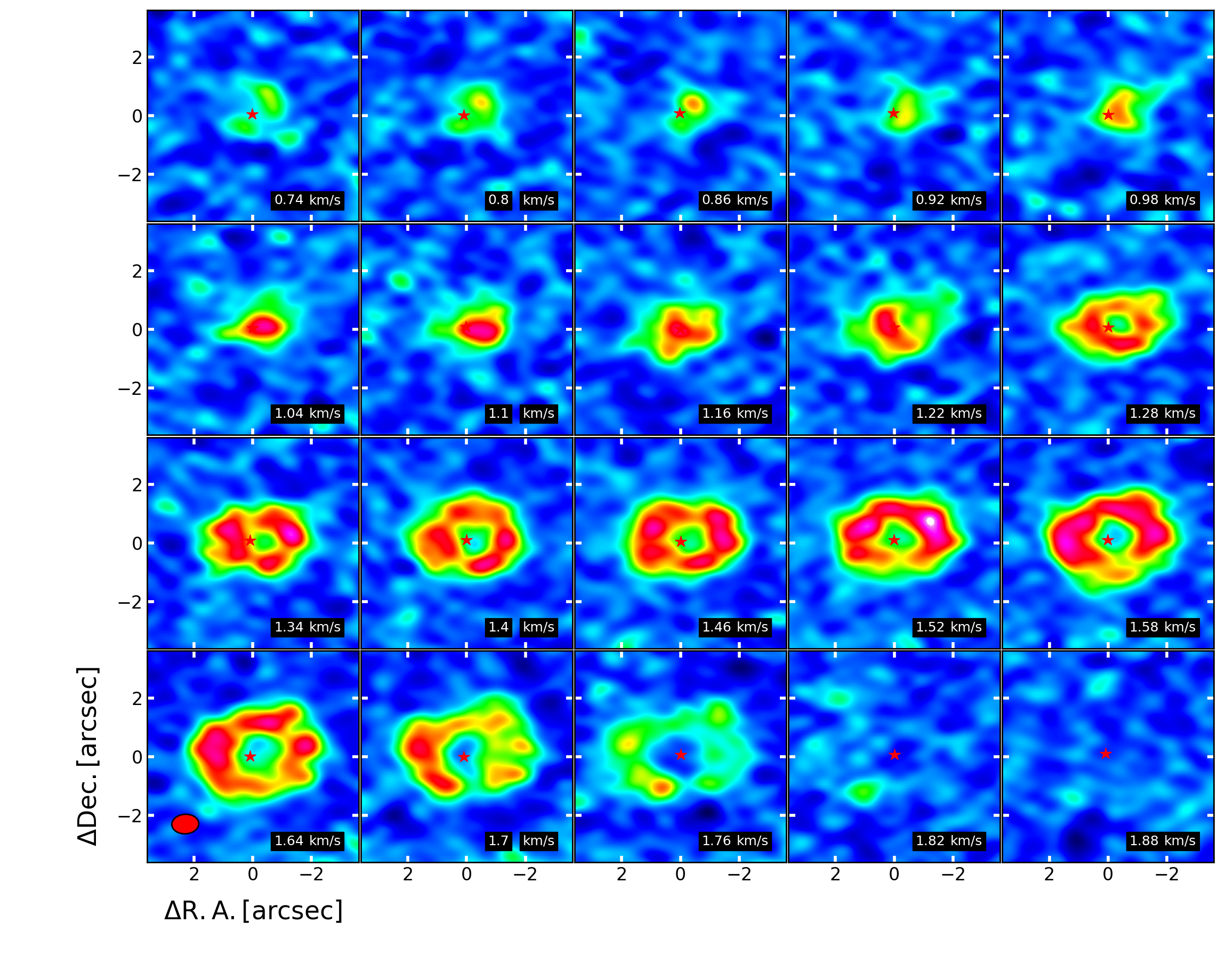}
\caption{Channel maps of the $^{12}$CO 230.53 GHz line around the location of the near-IR source SSTc2d J163134.1 (marked with a red star sign).  The $^{12}$CO channels at velocities larger than 1.9 km s$^{-1}$ are severely absorbed by colder gas in the foreground.} \label{Fig:ALMA_channel}
\end{figure*}

SSTc2d J163134.1 was observed as part of the ``Ophiuchus Disk Survey Employing ALMA'' (ODISEA) program (Project ID: 2016.1.00545.S PI: L. Cieza). ALMA Band 6 (1.3 mm) observations were performed on 2018 April 27th and August 22, during Cycle 5 using the C43-3 configuration (15-500 m baselines). The ODISEA sample was split into two Science Goals: the Class I  objects and the brighter Class II sources (K $<$ 10 mag)  were observed at resolution of 0.25$^{''}$ and sensitivity of 0.15 mJy beam$^{-1}$ continuum rms, while the fainter Class II  objects  (K $>$ 10 mag)  were observed at resolution of 0.8$^{''}$ and sensitivity of 0.25 mJy Beam$^{-1}$ continuum rms.  2MASS J163134 is part of the faint Class II sample. The total integration time was 54 sec on the science source with a field of view of 26$^{''}$. The correlator was set with four spectral windows in dual polarization mode, and configured to observe 60 MHz bands centered on the molecular transitions $^{13}$CO (J = 2 - 1) and C$^{18}$O (J = 2 - 1), and a 118 MHz band centered on $^{12}$CO (J = 2 - 1) line. The continuum spectral windows were centered on 233.02 and 218.02 GHz with bandwidths of 2.00 GHz each. The raw data were manually reduced by the North American ALMA Regional Center staff using CASA version 4.5.0. The flux density calibrator for the two execution blocks were J1517-2422 and J1625-2527, while the phase calibrator was J1517-2422 and the band-pass calibrator was J1733-1304. The corrected visibilities were imaged using the CLEAN algorithm after calibration of the bandpass, gain in amplitude and phase. The spatial resolution of the final image is 0.8 $\times$ 0.78 arcsec, with a PA of -88 deg., corresponding to 110 $\times$ 100 au at a distance of 139 pc, with a rms of 0.25 mJy beams$^{-1}$. Interestingly, ALMA  fails to detect continuum emission at 1.3 mm and $^{13}$CO and C$^{18}$O emission, while the $^{12}$CO line was imaged with Briggs weighting at a robust parameter of 0.5.  The Briggs weighting with a robustness of 0.5 was applied in both continuum and line imaging to obtain the optimal combination of resolution and image fidelity. Figure \ref{Fig:ALMA_channel} displays the final $^{12}$CO channel maps at 0.06 km s$^{-1}$ resolution between 0.74 and 1.88 kms$^{-1}$. The star symbol indicates the location of the IR source SSTc2d J163134.1, which is within $<$ 1.0$^{''}$ of the center of the $^{12}$CO bubble.

\subsection{KMOS}
\label{Sec:KMOS}

\begin{figure*}[ht!]
\epsscale{0.85}
\plotone{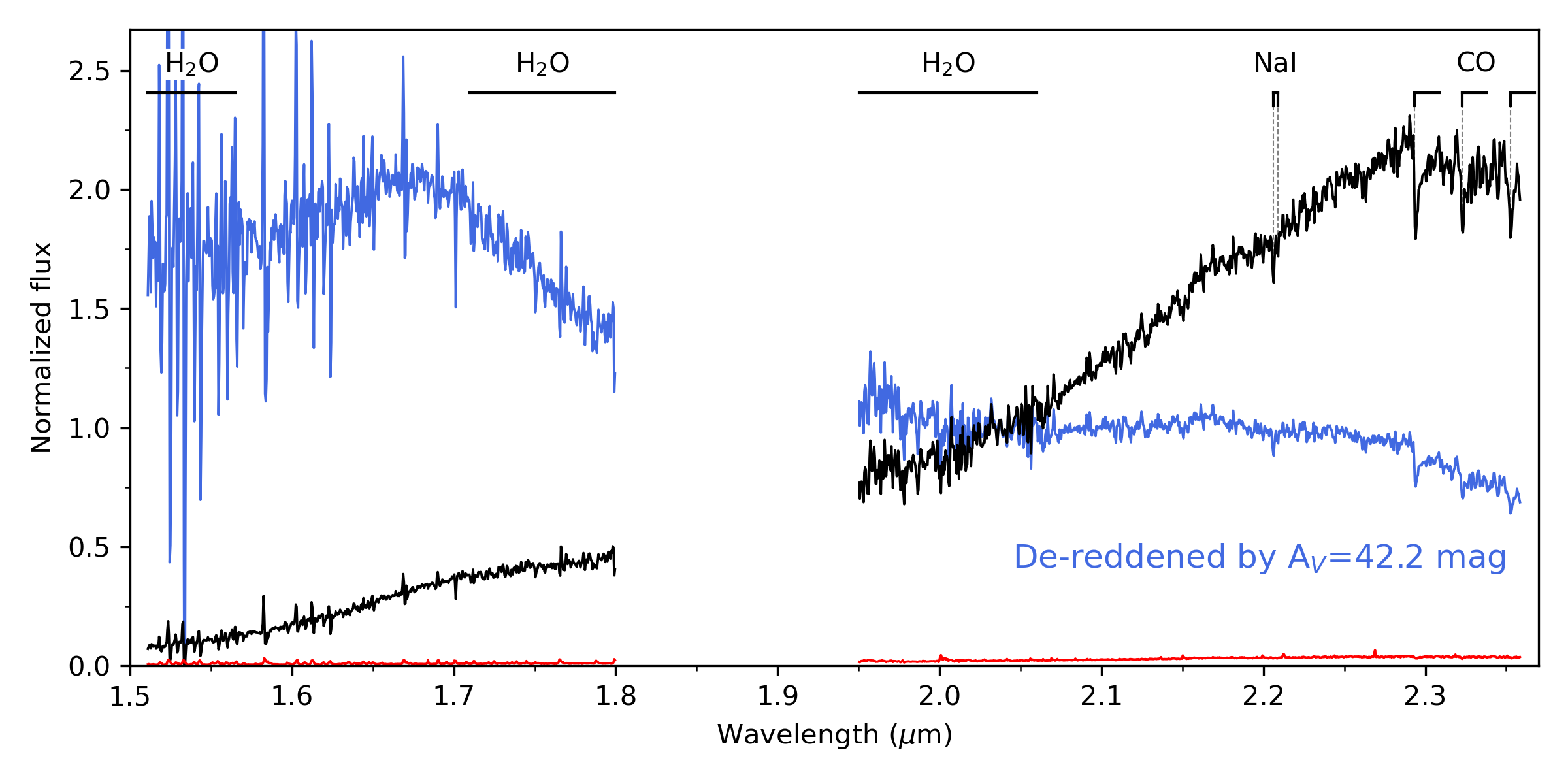}
\caption{The KMOS spectrum of SSTc2d J163134.1 and its associated noise are shown in black and red, respectively. The most relevant features present in spectra of very low mass objects are marked. The P-Cygni-like features in the H-band are due to imperfections in sky subtraction. The near-IR spectrum of SSTc2d J163134.1 from VLT-KMOS seen under 42 mag of visual extinction is displayed in blue. A more detailed description can be found in Sec. \ref{Sec:Spectroscopy}. The final KMOS spectrum of SSTc2d J163134.1 is available online.} \label{Fig:Kmos_spectrum}
\end{figure*}

In order to constrain the effective temperature and luminosity class of SSTc2d J163134.1, which is crucial to better understanding the nature of the enigmatic $^{12}$CO ``bubble'' emission, we obtained near-infrared spectroscopy of the stellar object using the K-band Multi-Object Spectrograph (KMOS) \citep{Sharples2013} at the ESO's Very Large Telescope (VLT), under ESO program ID 107.22RQ.001. KMOS performs Integral Field Spectroscopy (IFU) using 24 arms, each with a squared field of view (FoV) of 2.8$^{''}$ $\times$ 2.8$^{''}$, and the total FoV of the instrument of 7.2$^{'}$ in diameter. KMOS was operated in service mode using a nod to sky (ABAB) configuration. Two arms have been employed, with the sky offset defined so that in position A one of the arms observes the target and the other one a sky. At the B position, the first arm is allocated to a sky position and the second one to the target. With this strategy, we achieve (1) that the target is observed continuously throughout the observations, and (2) the optimal sky subtraction which requires observations of the target and the sky using the same IFU. Between each exposure, a small dither offset was applied. The Detector Integration Time (DIT) was 130 s and a total of 49 exposures were obtained, using the HK band filter ($\sim$1.5-2.4 $\micron$), providing a mean resolving power of $\sim$1500. Telluric standard stars (spectral B-type) at a similar air-mass have been observed immediately before or after the science observations.

The data reduction was performed using the KMOS pipeline \citep[$\rm SPARK$][]{Davies2013} in the ESO Reflex automated data reduction environment \citep{Freudling2013}. The Reflex environment allows to carry out data reductions in an almost automatic way, with several parameters that can be adjusted. The pipeline produces calibrated 3D cubes for each individual exposure, by performing flat field correction, wavelength calibration, sky subtraction, and telluric correction. The telluric correction is performed with the help of the Molecfit tool \citep{Smette2015, Kausch2015}, which fits synthetic transmission spectra to the astronomical data,  in our case the telluric standard stars. We extracted the spectra using a 2D gaussian mask and median combined them. The noise at each wavelength was derived as:

\begin{eqnarray*}
N = \frac{\sigma }{\sqrt{n}}
\end{eqnarray*}

where N is the noise, $\sigma$ is the standard deviation of all the exposures at that wavelength, and n is the number of exposures. The final spectrum is shown in Fig. \ref{Fig:Kmos_spectrum}, together with the most prominent features present in the spectrum of low-mass stars and brown dwarfs. More details can be found in Section \ref{Sec:Spectroscopy}.

\section{Results and Analysis} 
\label{Sec:Analysis}

\begin{figure*}[ht!]
\epsscale{0.85}
\gridline{\fig{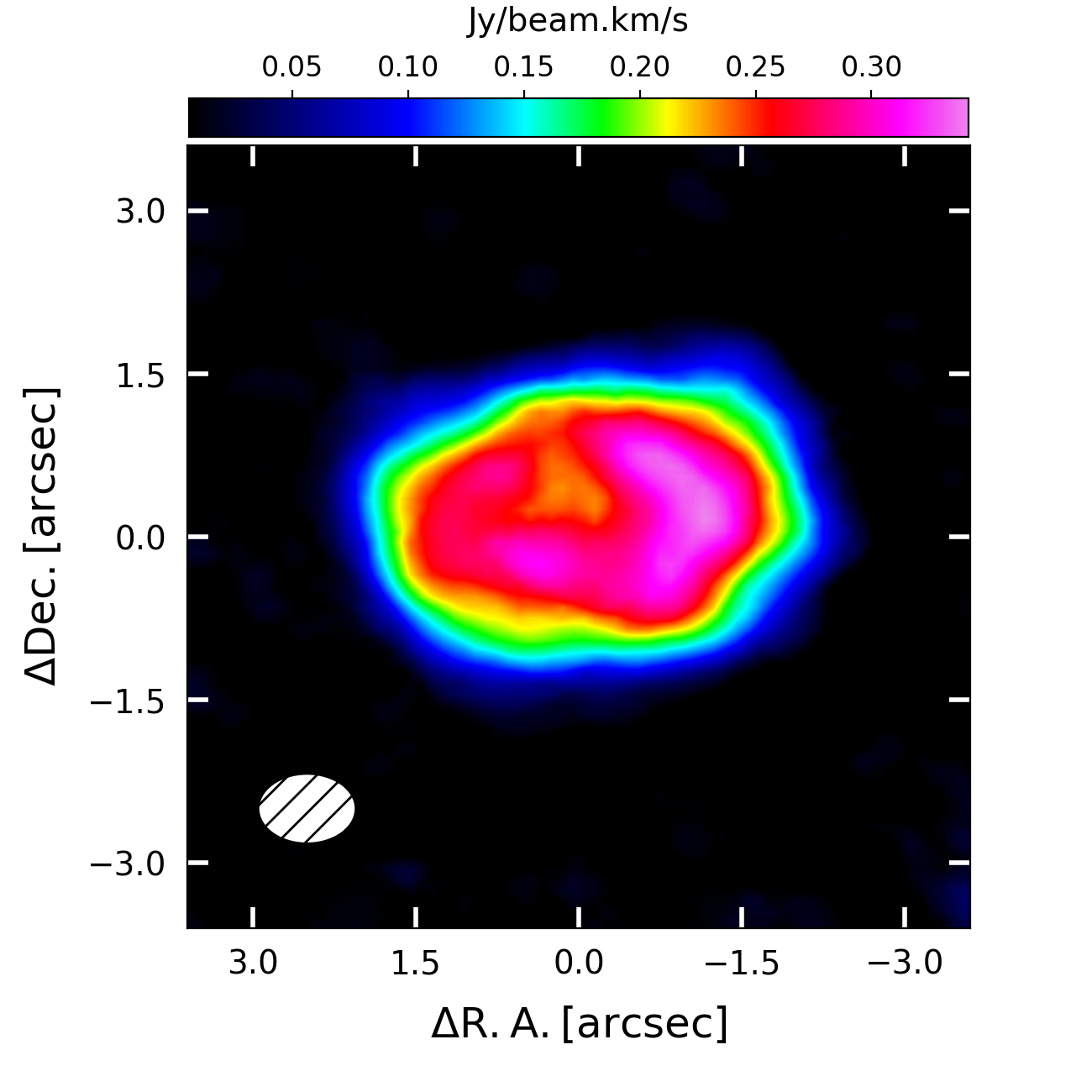}{0.33\textwidth}{(a)}
         \fig{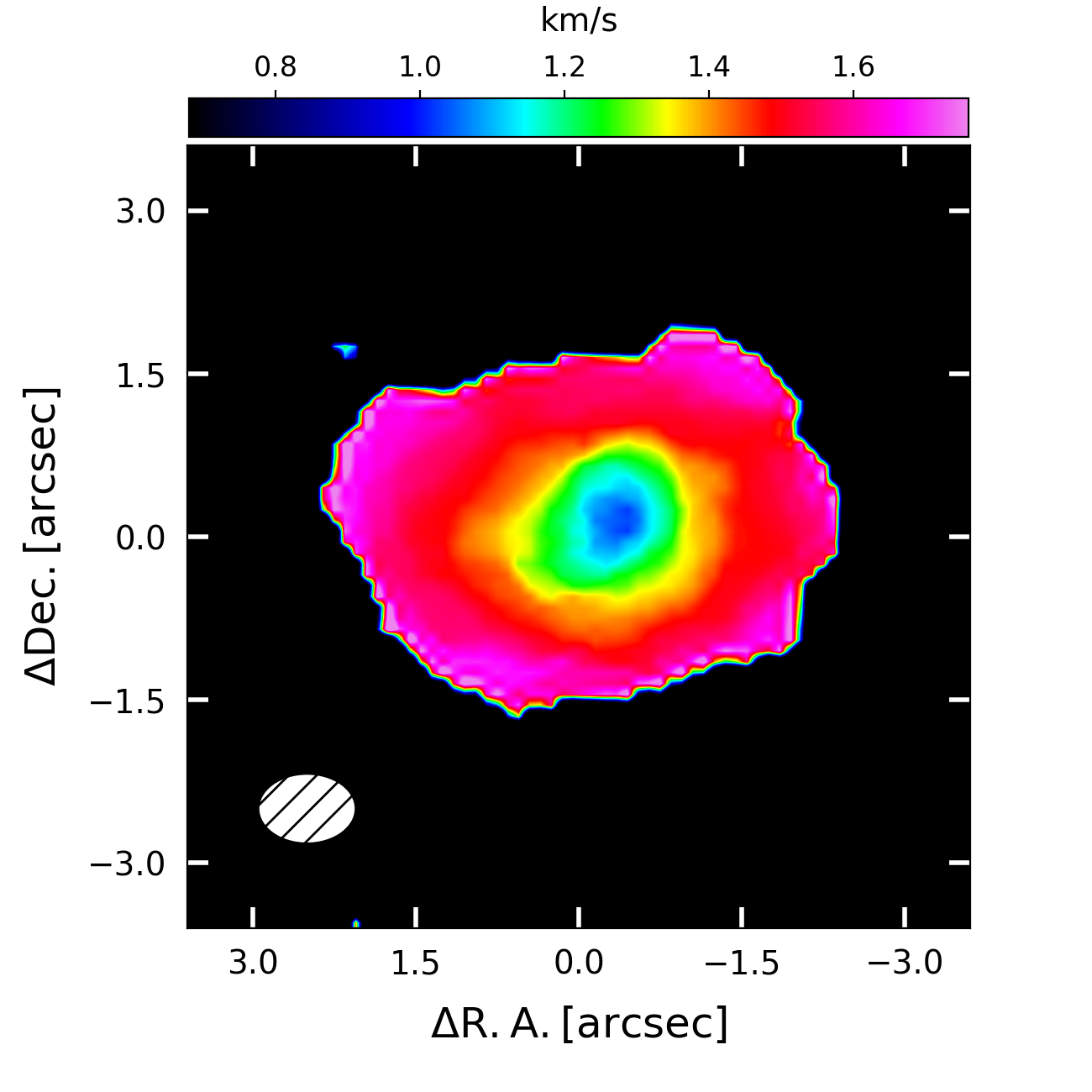}{0.33\textwidth}{(b)}
         \fig{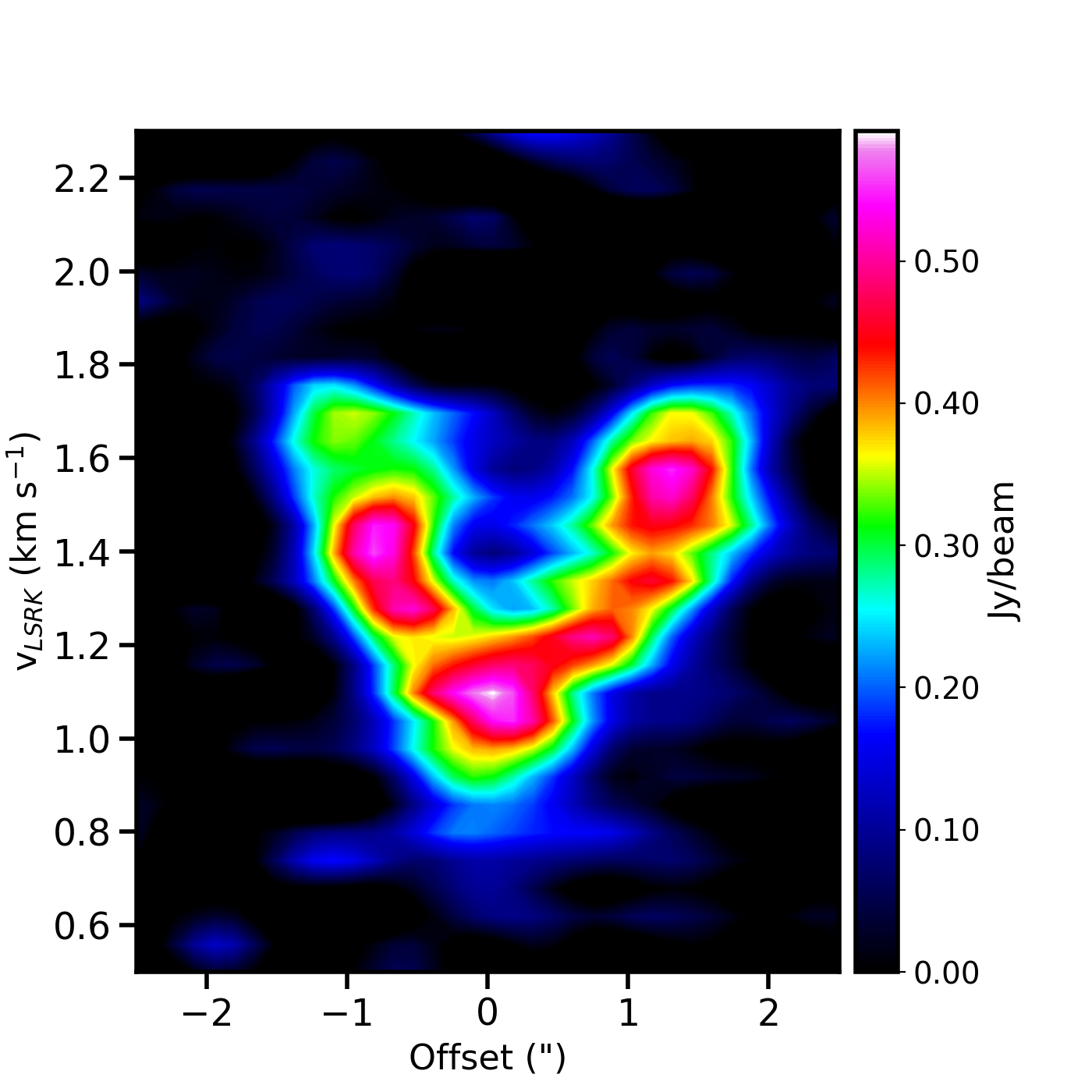}{0.33\textwidth}{(c)}}
\caption{ALMA blue-shifted $^{12}$CO emission line towards SSTc2d J163134.1. a) Integrated intensity maps (Moment-0). b) Intensity weighted mean velocity (Moment-1). The line integrated emission of the $^{12}$CO line was performed in the range from 0.7 to 1.88 km s$^{-1}$. In each panel, the corresponding beam size is shown at the bottom left. c). Position-velocity diagram along a P.A. of 45 deg. of the observed $^{12}$CO spherical emission shape. The position-velocity diagram demonstrates that we are detecting the blue-shifted portion of an expanding or collapsing sphere, see Sec. \ref{Sec:Model}.} \label{Fig:Moment}
\end{figure*}

\subsection{$^{12}$CO Line Emission}

Out of a sample of almost 300 YSOs in the ODISEA project, SSTc2d J163134.1  stands out as a very bright ellipse of $^{12}$CO gas 3$^{''}$ $\times$ 4$^{''}$  (Fig. \ref{Fig:ALMA_channel}) without a broadband 1.3 mm counterpart. Figures  \ref{Fig:Moment}a and \ref{Fig:Moment}b show the velocity integrated map (moment 0) together with the  velocity-weighted intensity map (moment 1) of the $^{12}$CO(J = 2-1) line at 230.53 GHz integrated over a range between 0.7 and 1.88 km s$^{-1}$, respectively. The moment-1 map of the $^{12}$CO line shows the characteristic signatures of an expanding or collapsing shell of gas: a radially symmetric velocity field with the highest approaching velocity at the center (Fig. \ref{Fig:Moment}). The position-velocity (PV) diagram of the $^{12}$CO line across the gas clump at a PA of $\sim$45 deg. also shows the blue-shifted ``V'' shape signature of outflowing/infalling motion (Fig. \ref{Fig:Moment}c). The redshifted emission is severely affected by self-absorption by colder foreground material, preventing us from  detecting the inverted``v" shape expected for material at the corresponding velocity.

\subsection{Derivation of the spectral type and extinction}
\label{Sec:Spectroscopy}

\begin{figure*}[ht!]
\epsscale{0.85}
\plotone{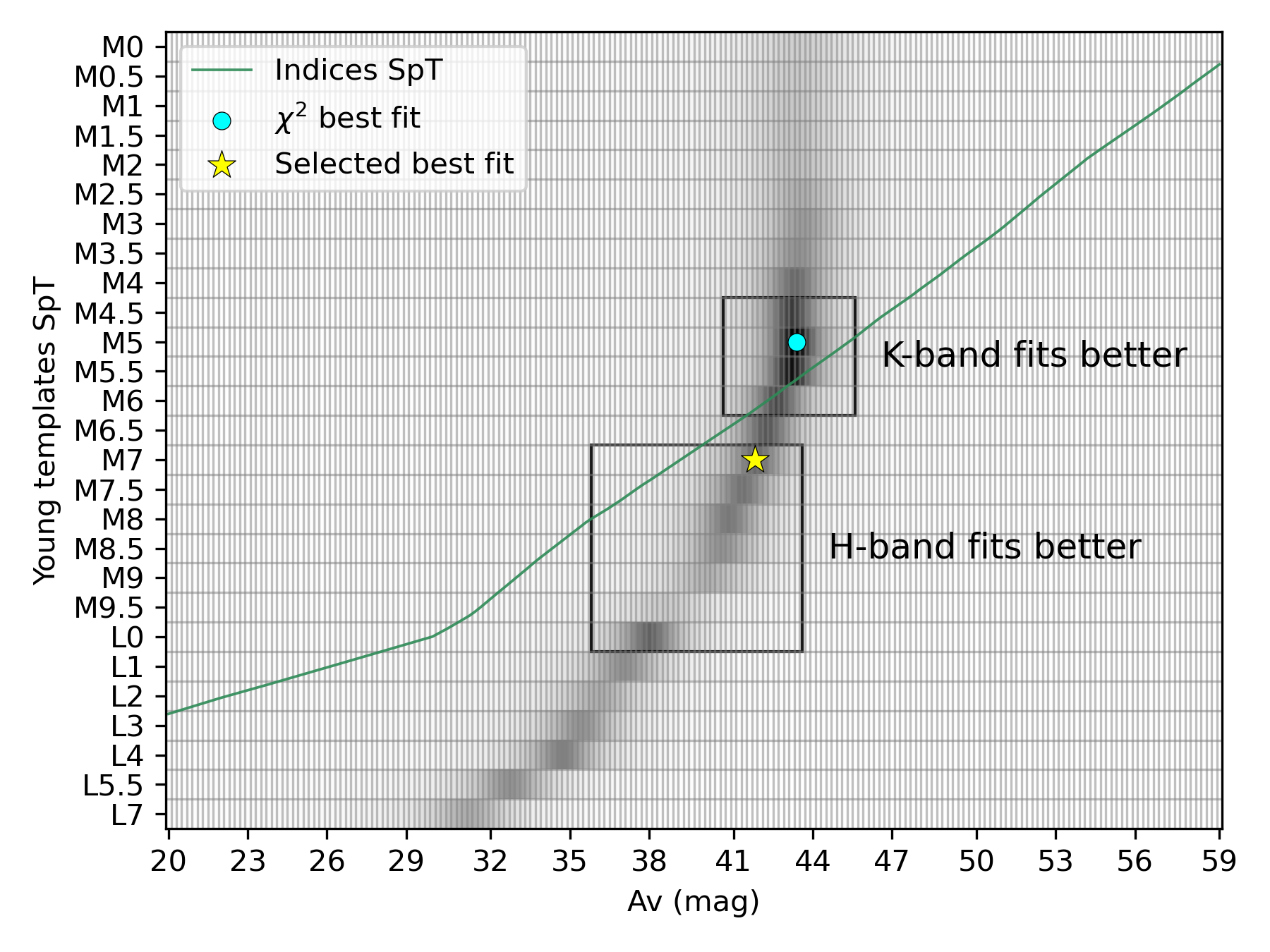}
\caption{The probability map for different SpT-Av combinations, where darker regions represent lower values of $\chi^{2}$. The full spectrum of SSTc2d J163134.1 was compared to the young dwarf spectral templates from \citet{Luhman2017}. The green line indicates the spectral type derived from a combination of various spectral indices (see Sec. \ref{Sec:Spectroscopy}) for each value of Av.} \label{Fig:Indices}
\end{figure*}

We derive both the spectral type (SpT) and extinction simultaneously by direct comparison between the observed IR spectrum (see Sec. \ref{Sec:Data}) with young spectral templates from \citet[][age $\leq$ 10 Myr]{Luhman2017} and with field spectral templates from \citet{Kirkpatrick2010}. Considering that the K-band spectrum shows the CO band heads of 2.29 $\micron$ in absorption which are typical of late-type Class II/III YSOs, as well as of evolved red stars \citep[e.g post-Asymptotic Giant Branch (AGB) stars;][]{Wallace1997}, we also compare the object spectrum with giant star spectra from the IRTF spectral library \citep{Cushing2005, Rayner2009}. The comparative process was performed with each spectral template for a wide grid of extinction values Av=0-60 mag with a step of 0.2 mag, and the extinction law from \citet{Cardelli1989} with Rv=3.1. The comparison is made for the H and K bands neglecting the telluric region in between them (1.8-1.95 \micron). The goodness of fit of each comparison is evaluated using the $\chi^{2}$ minimization:

\begin{eqnarray*}
    \chi^{2} = \frac{1}{n-m}\sum_{i}^{n}\frac{O_{i}-T_{i}}{\sigma ^{2}}
\end{eqnarray*}

where $\rm O$ is the object spectrum, $\rm T$ the template spectrum, $\sigma$ is the noise of the observed spectrum, $\rm n$ the number of data points, and $\rm m$ the number of fitted parameters ($\rm m=2$).

The results for the set of young object templates are shown Fig. \ref{Fig:Indices} in the form of a probability map, where the darkest regions represent lower $\chi^{2}$. Cool dwarfs spectra typically show some degree of degeneracy between the spectral type and the extinction \citep{Luhman2017}, which is also evident in this case, due to the large amount of extinction through which we see the object. When fitting the two bands individually, we find that the earlier SpTs (M5-M7) provide a better fit in the K-band, while the H-band portion is better represented by SpTs between M7 and L0. Thus, the $\chi^{2}$ is minimized at M5 with Av=43.8 mag, which follows the K-band morphology of the target spectrum but fails at reproducing the beginning of the H-band ($\lambda<$1.66 $\mu m$). At later SpTs (down to L1) we observe that the H-band shape is better matched while the K-band fit worsens. In any case, the derived extinction is in excellent agreement with the value derived from the Herschel column density maps (see Fig. \ref{Fig:Density}). The final results for the set of young object templates are presented in Appendix \ref{App:Spectra}. Figure \ref{Fig:Indices} also shows the SpT derived using spectral indices, for each value of the extinction grid (green line). The spectrum is first de-reddened, and the spectral type is then derived as a mean of the values obtained by 4 different spectral indices: WK\citep{Lucas2006}, H$_{2}$O$_{2}$ \citep{Slesnick2004}, sH$_{2}$OK \citep{Testi2001}, and TLI-K \citep{Almendros2022}. All indices are defined in the K-band, which is likely the reason why the $\chi^{2}$ map intersects with the green line at $\sim$M6, where the templates fit the K-band better. The target spectrum is therefore constrained to M5-L0 and will be further discussed in Sec. \ref{Sec:ClassII}.

\begin{figure*}[ht!]
\epsscale{0.85}
\plotone{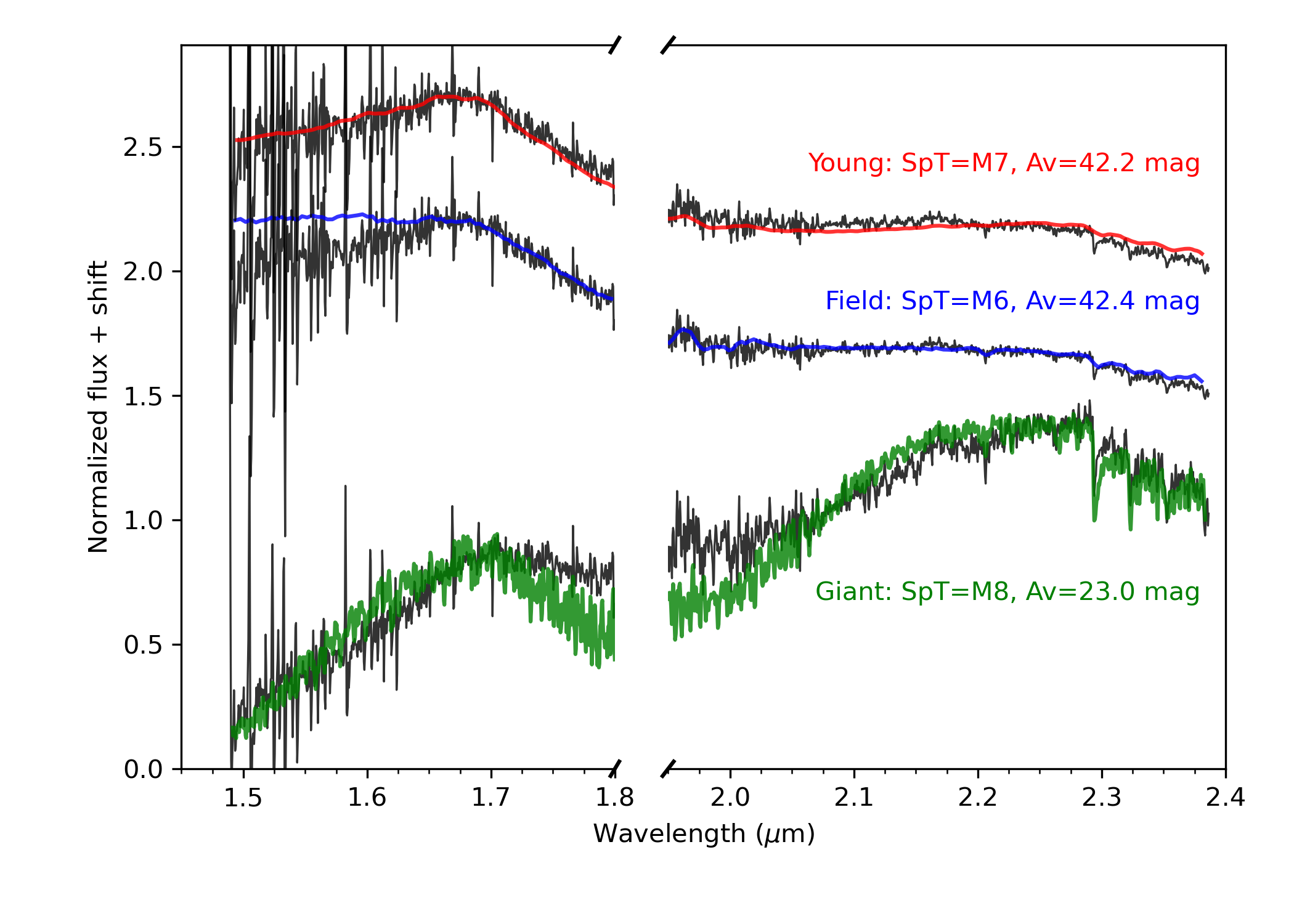}
\caption{Best-fit templates for each of the object classes: young \citep[red;][]{Luhman2017}, field \citep[blue;][]{Kirkpatrick2010} and giants (green; IRTF spectral library). In black, we show the spectrum of SSTc2d J163134.1, de-reddend by the corresponding extinction. The spectra have been normalized at 1.66 $\micron$, and shifted for clarity.} \label{Fig:Spectra_Best}
\end{figure*}

\begin{figure*}[ht!]
\epsscale{0.85}
\gridline{\fig{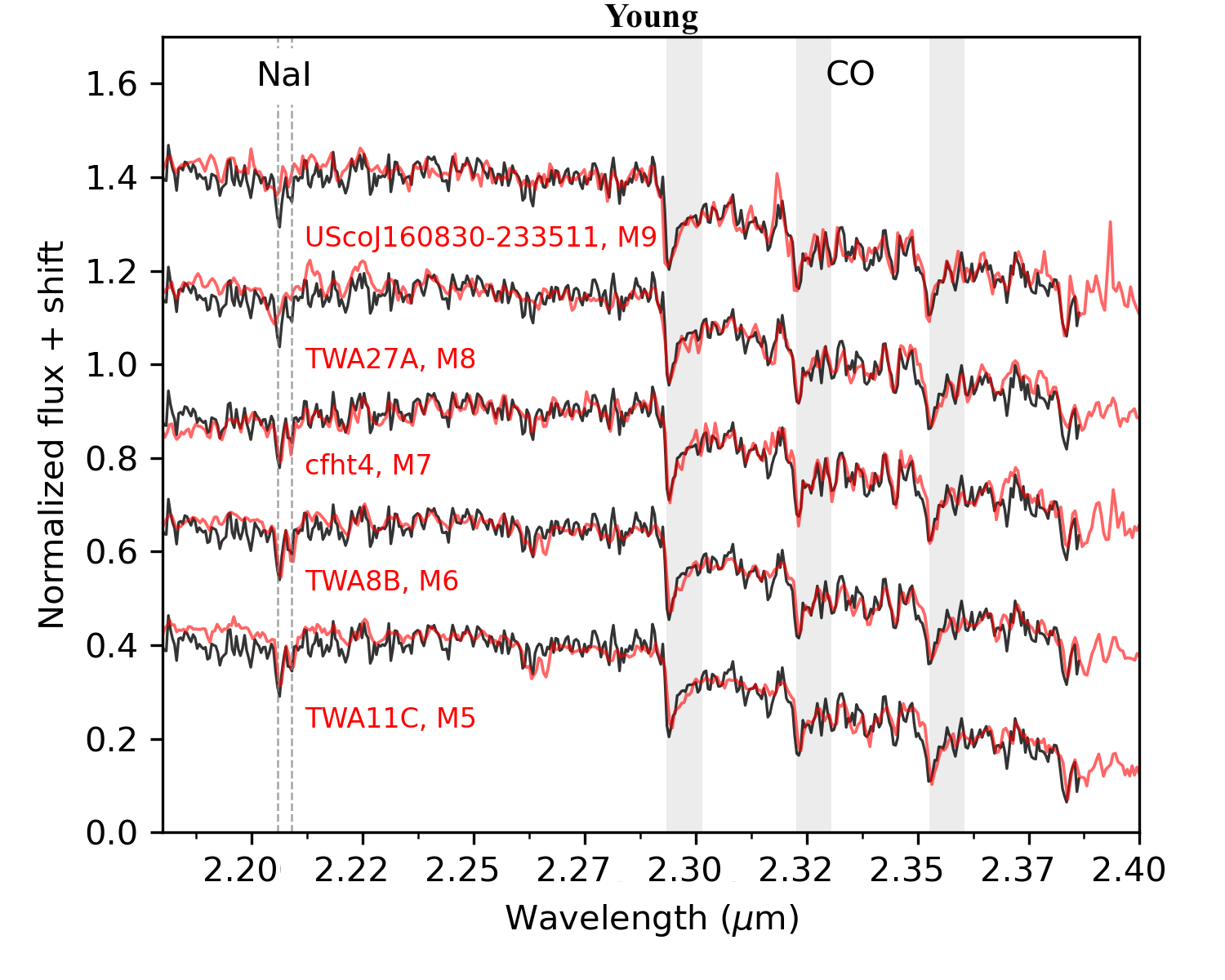}{0.33\textwidth}{(a)}
         \fig{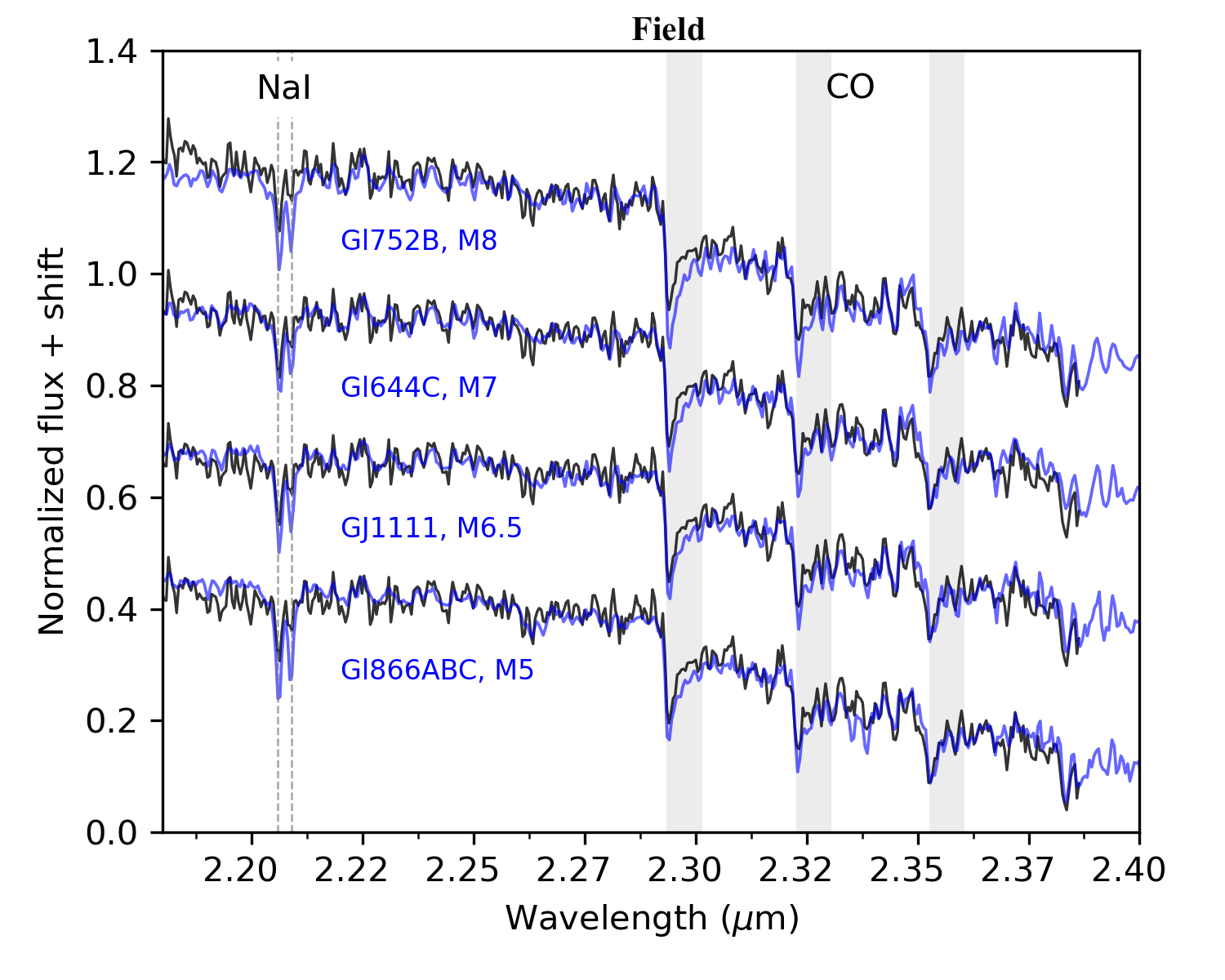}{0.33\textwidth}{(b)}
         \fig{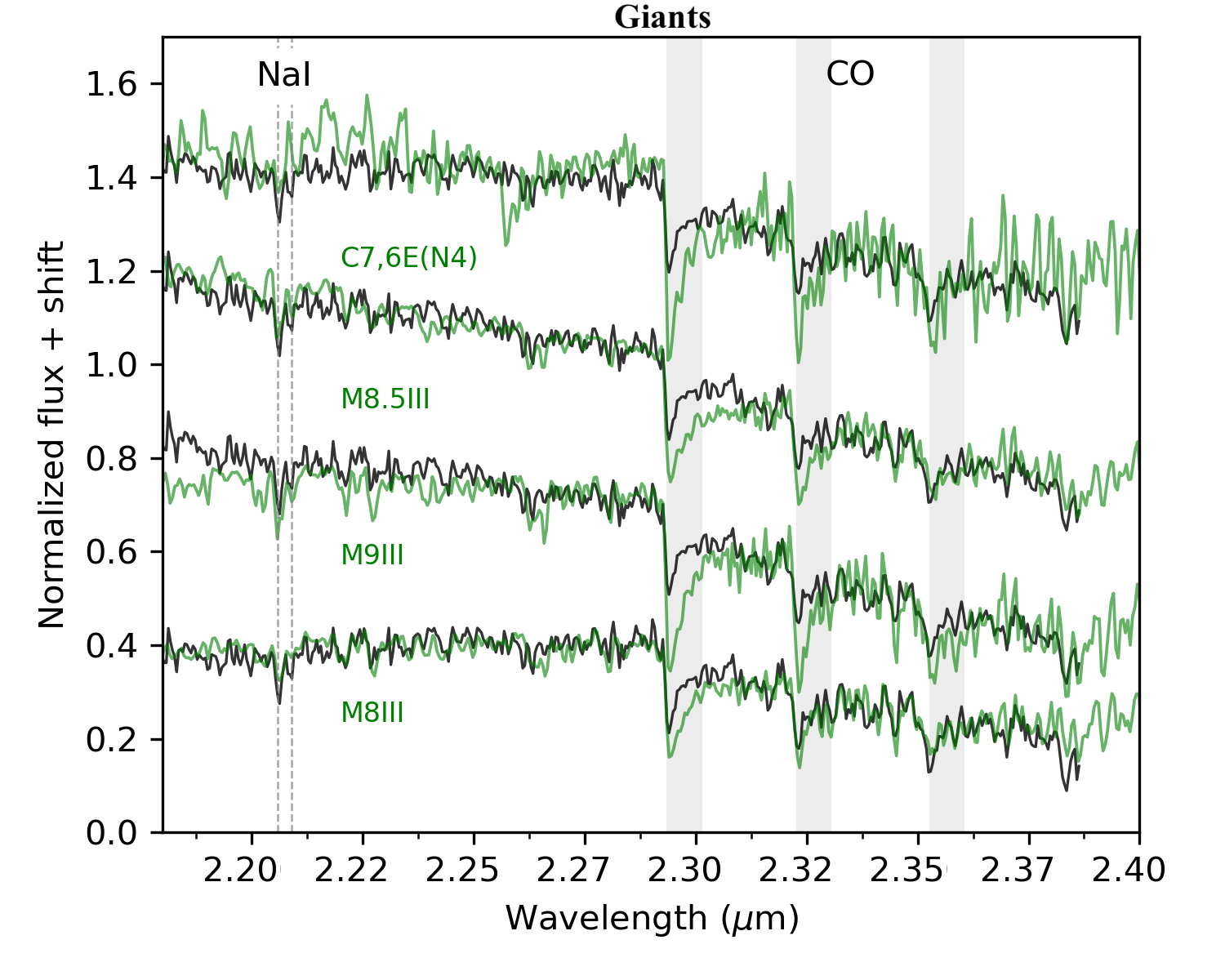}{0.33\textwidth}{(c)}}
\caption{A zoom into the K-band region of SSTc2d J163134.1 (black line), compared to the best-fit spectra for each class at a comparable spectral resolution; a) young dwarfs, b) field dwarfs, and c) giant stars. We highlight the position of the NaI doblet and the most important CO band-heads. The spectra have been normalized at 1.66 $\micron$, and shifted for clarity.} \label{Fig:Spectra_Best2}
\end{figure*}

\subsubsection{SSTc2d J163134.1: A late-type Class II/III YSO or an evolved red star?}
\label{Sec:ClassII}

While the template fitting provides a robust criteria for the classification of SSTc2d J163134.1 as an M-type star, the distinction between a late-type Class II/III YSO and an evolved red star also needs to be considered as they are totally  distinct in evolutionary phase. To that end, there are several gravity-sensitive features that can be used to further constrain the nature of the object. Figure \ref{Fig:Spectra_Best} displays the best-fit spectra belonging to the young dwarf, field dwarf, and giant class, along with the object spectrum de-redenned by the corresponding Av. First, the overall sharp triangular shape of the H-band in young objects (caused by the water absorption), as opposed to a rounder form in old field objects \citep{Cushing2005, Kirkpatrick2006}. Field dwarf templates do not resemble well the shape of the first half of the H-band, which is indicative of the young nature of the object \citep{Lucas2001}. Furthermore, the object spectrum is generally not well represented by the giant spectra, which also provide the best fits at significantly lower extinction values than those expected from the Herschel column density map (see Fig. \ref{Fig:Density}).

In addition to the broad spectral features, we can look at some specific line regions. In the K-band portion of the KMOS spectrum, we can resolve the NaI doublet at $\sim$2.21 $\micron$, and the CO bandheads longwards of 2.29 $\micron$ (Fig. \ref{Fig:Spectra_Best2}). While the alkali-metal lines in the near-infrared in general exhibit a gravity-sensitive behaviour, the NaI doublet at $\sim$2.21 $\micron$, however, only weakly correlates with surface gravity \citep{Gorlova2003}. The CO bandheads, on the other hand, can be used to distinguish between dwarfs and giants, as the former show significantly weaker CO absorption at SpTs $>$ M0 \citep{Baldwin1973, Foster2000, Gorlova2003}. In order to make use of these features, we need to compare the spectrum with higher resolution spectra (R $>$ 1000) than those of the spectral templates. For that, we use young dwarf spectra from \citet{Luhman2007, Allers2013, Bonnefoy2014, Manjavacas2014} and \citet{Lodieu2008}, as well as the field dwarfs from \citet{McLean2003, Cushing2005} and \citet{Rayner2009}. We follow the same comparison method introduced previously but restricted to the region encompassed by these features (2.19-2.38 $\mu m$). Figures \ref{Fig:Spectra_Best2}a and \ref{Fig:Spectra_Best2}b display the M5-M9 and M5-M8 sequence of best-fit moderate resolution young and field spectra, respectively. We observe that the M5-M7 young spectra reproduce nicely the depth and width of the NaI doublet and CO bandheads. The NaI doublet of the moderate field spectra is consistently stronger than in our observed KMOS spectrum, and the first two CO bandheads are not well reproduced either. Figure \ref{Fig:Spectra_Best2}c shows the same, but for the four best fit giants, and we observe that the CO bandheads are significantly stronger than the observed KMOS spectrum and the NaI doublet is not well matched either. The final results for the set of field dwarf and giant class templates are presented in Appendix \ref{App:Spectra}.

To conclude, the near-infrared spectrum of SSTc2d J163134.1 is best represented by the young dwarf spectral templates, with several gravity-sensitive features speaking against the giant nature. Additionally, the strength of the NaI doublet allows us to constrain the SpT of the target to M5-M7, where M7 is the SpT that best reproduces the shape of the H-band. Taking all of this into account, we consider the target spectrum is best represented by a young BD with SpT M7 and Av=42.2 mag. Figure \ref{Fig:Kmos_spectrum} displays the KMOS spectrum of SSTc2d J163134.1 seen under an Av of $\sim$ 42 mag. Nevertheless, as we will discuss in Section \ref{Sec:Scenario}, additional kinematic studies are needed to further constrain the location and provide additional confidence that SSTc2d J163134.1 is a true YSO member of the Ophiuchus complex dark clouds \citep[i.e. 2-6 Myrs][]{Esplin2020}.

\subsection{Spectral Energy Distribution}

\begin{figure*}[ht!]
\epsscale{0.6}
\plotone{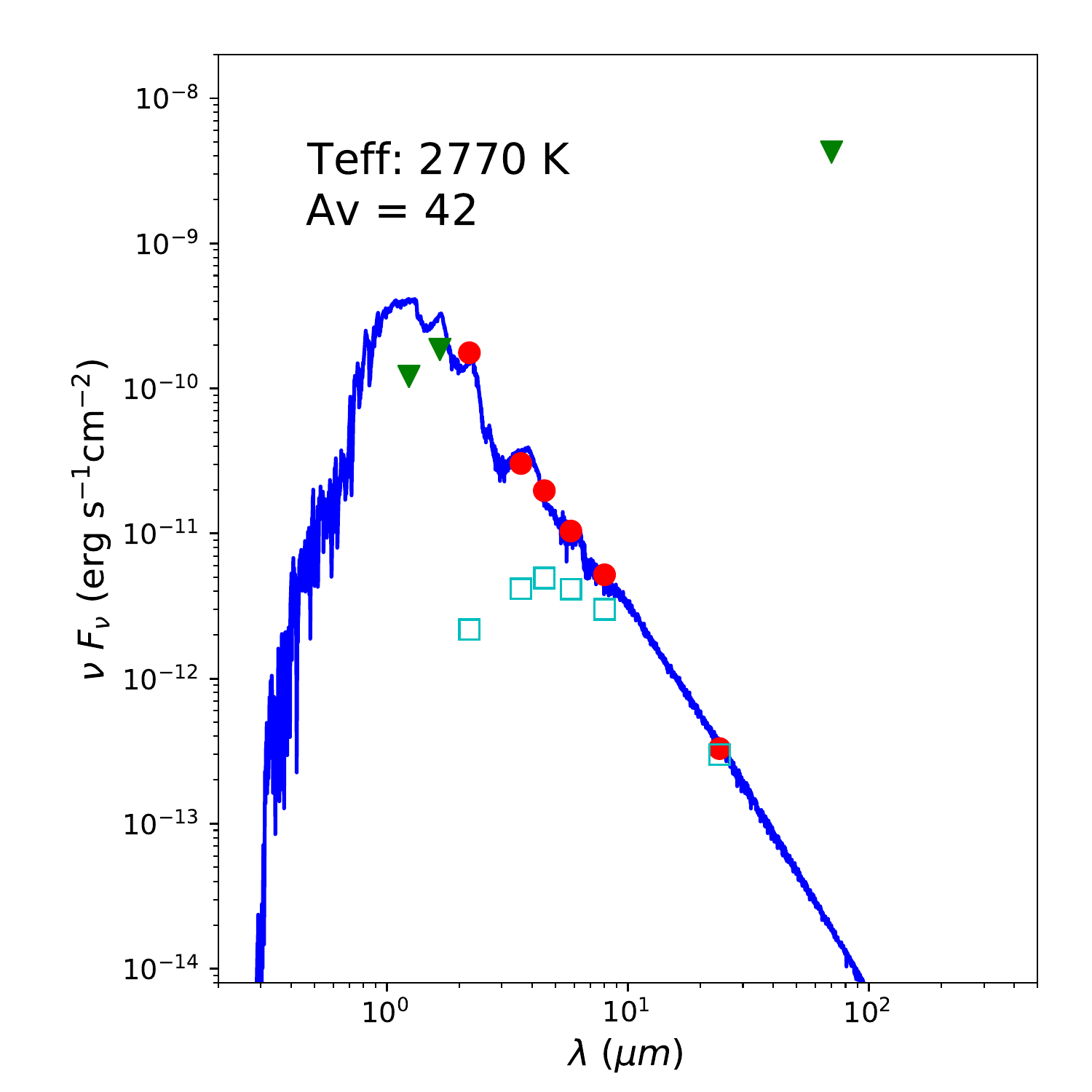}
\caption{Spectral energy distribution for SSTc2d J163134.1 built with   photometric data acquired from the literature between 1.6 and 24 $\micron$ \citep{Cutri2003, Evans2009}. Cyan squares represent the observed IR photometry before correcting for extinction. Red dots represent the IR photometry corrected for an extinction of Av $\sim$ 42 adopting the extinction law from \citet{Cardelli1989} with Rv = 3.1. Green triangles show IR photometry upper limits. The blue line is the BT-settl spectrum model-- according to the spectral type (see Sec \ref{Sec:Spectroscopy})-- multiplied by the corresponding M$_{d}$ factor. Accordingly, this fitting is consistent with a bare (diskless) stellar template of 2770 K \citep{Baraffe2015} seen under an Av $\sim$42 mag. \label{Fig:SED}}
\end{figure*}

As an aid for interpreting the evolutinary phase of SSTc2d J163134.1, we built its SED with photometry measurements compiled from the literature - where available - thus collecting the near-infrared J, H, and K photometry from 2MASS and the Spitzer IRAC and MIPS photometry \citep{Cutri2003, Evans2009}. Figure \ref{Fig:SED} displays the resulting SED of SSTc2d J163134.1 adopting the spectroscopic parameters derived in Sec. \ref{Sec:Spectroscopy}. To reproduce the stellar contribution to the SED, we used the BT-Settl model spectra \citep{Baraffe2015}, where the adopted SpT M7 is converted into its effective temperature (T$\rm _{eff}$ = 2770 K) using the relation presented in \citet{Herczeg2014}. The observed SED has been dereddened using the visual extinction estimated in Sec. \ref{Sec:Spectroscopy} ($A_{V}$ = 42 mag.) by adopting the extinction law of \citet{Cardelli1989} and assuming $\rm R$ = $\frac{A_{V}}{E_{(B-V)}} = 3.1$. In addition, the SED is normalized at the distance of the Ophiuchus complex dark cloud L1709 \citep[139 pc; ][]{Esplin2020}. We find that when the 1.6-24$\micron$ SED is corrected for the visual extinction derived from the 1.6-2.4 $\micron$ spectrum, the dereddened SED is consistent with a bare stellar photosphere (diskless) with a T$\rm _{eff}$ = 2770 K $\pm$ 100, as shown in Fig. \ref{Fig:SED}. Accordingly to the evolutionary models \citep{Baraffe2015} and assuming its location in the L1709 dark cloud, SSTc2d J163134.1 with an age of $\sim$ 2 Myr is an M7 star with a stellar mass of $\sim$0.05 M$_{\odot}$ and luminosity of $\sim$0.02 L$_{\odot}$.

\section{Possible Scenarios for the $^{12}$CO ``bubble" shape}
\label{Sec:Scenario}

\begin{figure*}[ht!]
\epsscale{0.9}
\plotone{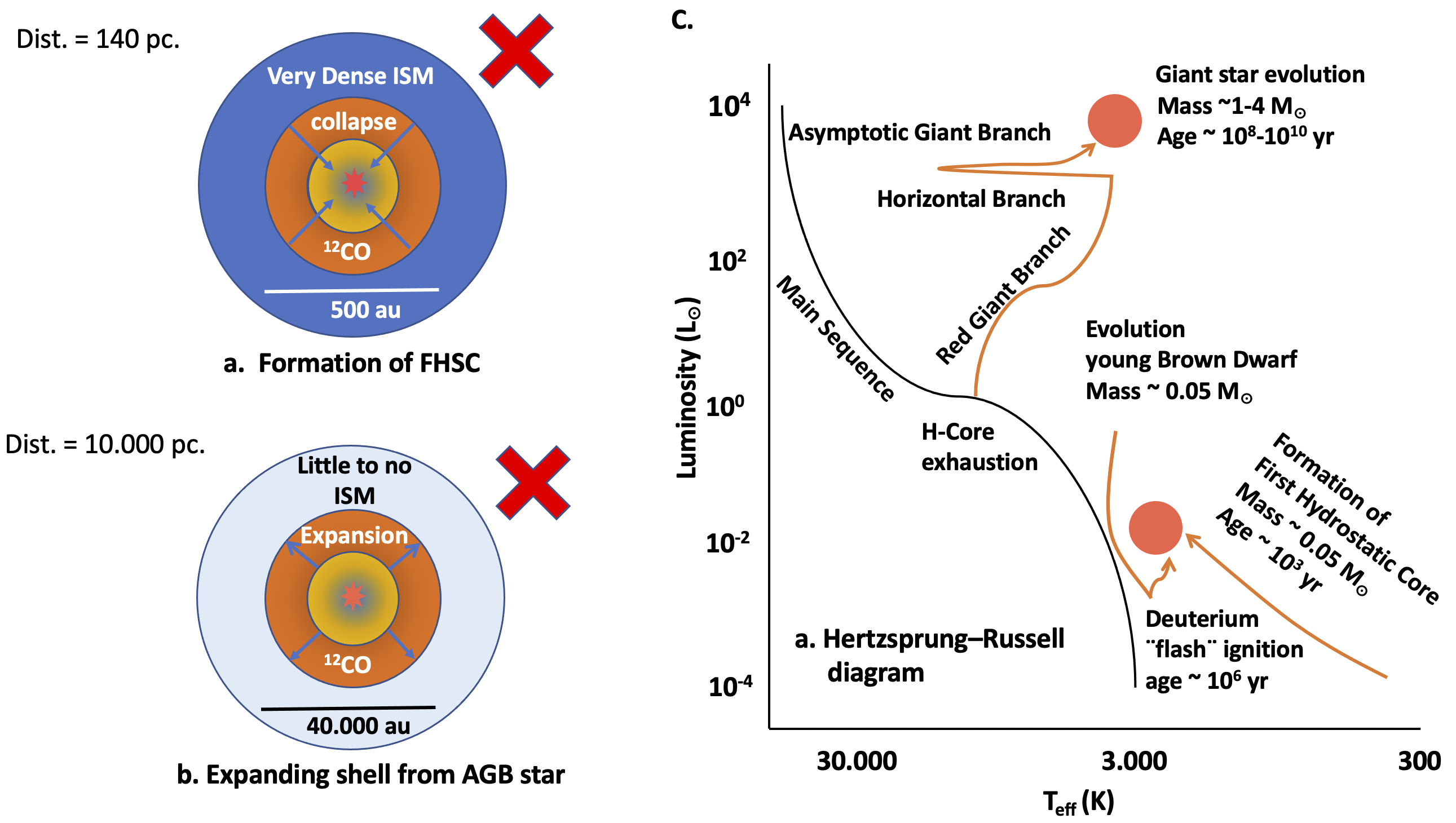}
\caption{Panels a. and b. show schematics of the two scenarios we have initially considered in order to describe the $^{12}$CO ``bubble" (see Sec. \ref{Sec:Scenario}). A First hydrostatic core (a.) can be ruled out based on the SED of the object and the $^{13}$CO upper limits, indicating moderate gas densities. An AGB star (b.) can also be ruled out based on the 3-D motion of the object, indicating that  SSTc2d J163134.1 belongs to the Ophiuchus molecular cloud at a distance of 139 pc, and thus should have sub-solar luminosity. Panel c. shows the approximate position in the HR diagram of the astrophysical objects considered. The orange arrows show possible evolutionary tracks. The orange rings correspond to the expanding/collapsing shell of $^{12}$CO gas detected by ALMA. An expanding shell from a BD remains the only viable explanation as further presented in Sec. \ref{Sec:Model}.} \label{Fig:Cartoon}
\end{figure*}

In order to disentangle the nature, location and physical characteristics of the detected $^{12}$CO ``bubble", and its connection to SSTc2d J163134.1, we initially considered two known astrophysical phenomena given all of the available evidence:
 
 \begin{enumerate}
    \item {The inside-out collapse of a dense molecular core in the Ophiuchus cloud.}
    \item {The mass loss of a giant star in the distant background (d $>$ 5-10 kpc).}
\end{enumerate}

These two scenarios are illustrated in Figure \ref{Fig:Cartoon} together with the approximate position in the Hertzsprung–Russell (HR) diagram of the astrophysical objects considered.

 \subsection{Scenario 1: The inside-out collapse of a dense molecular core in the Ophiuchus cloud.} In this scenario, the ALMA $^{12}$CO observations would trace the free-fall region of the collapse and the IR source would represent an extremely young object, possibly a \textit{first hydrostatic core}, a type of young stellar object that has not conclusively been observed to date \citep{Young2019}. Unfortunately, this scenario presents very significant downsides. In particular, the lack of ALMA continuum and the IR SED  (Figs. \ref{Fig:ALMA_channel} and \ref{Fig:SED}) are inconsistent with the properties of very young sources embedded in the molecular clouds, which are characterized by a SED steeply rising in the mid-IR and very strong dust mm emission, of the order of 1000 mJy at a distance of 150 pc \citep{Commercon2012}. This predicted flux is 3 orders of magnitude higher than our 1.3 mm upper limits ($\sim$4$\sigma$ limit of $\sim$1 mJy).

 \subsubsection{Infall or Outflow motion?}
 
 In addition to the lack of dust emission, the ALMA upper limits on $^{13}$CO emission towards SSTc2d J163134.1 are also inconsistent with infall motion in a very dense environment. Overall, the $^{12}$CO(2-1)/$^{13}$CO(2-1) line ratio towards molecular cores or protoplanetary disks is typically of $\sim$3 in velocity-integrated intensity, or even up to 2 in peak intensity. This is indeed the case for single-dish observations of the Ophiuchus complex, including the line of sight towards SSTc2d J163134.1, as can be concluded from the $\rm COMPLETE$ database where observations acquired at FCRAO are available \citep{Ridge2006, 12CO2011, 13CO2011}. These single-dish data towards SSTc2d J163134.1 show strong $^{12}$CO(1-0) and $^{13}$CO(1-0) emission, fairly Gaussian in shape and singly-peaked at VLSR = 2.6 km s$^{-1}$, with a FWHM of 2 km s$^{-1}$. However, this diffuse emission is largely filtered out by the ALMA interferometer, and it is offset by 1.5 km s$^{-1}$ from SSTc2d J163134.1. The lack of detectable $^{13}$CO(2-1) towards SSTc2d J163134.1 is therefore suggestive of low CO column densities, much lower than observed in dense molecular cores \citep{Bergin2007}.

 \begin{figure*}[ht!]
\epsscale{0.85}
\plotone{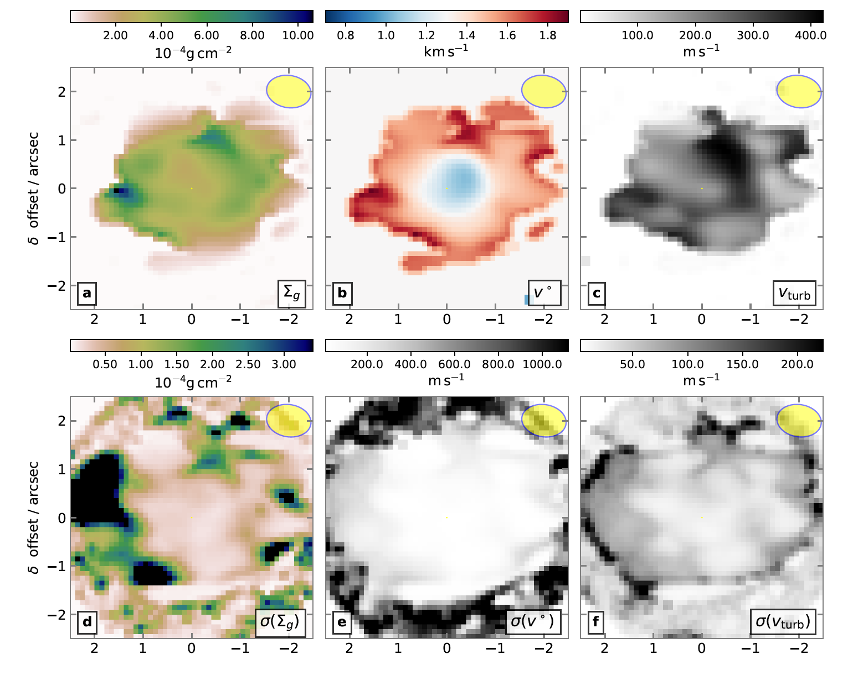}
\caption{Line diagnostics towards SSTc2d J163134.1 obtained by fixing the temperature $\rm T$ = 30 K. We show the total molecular mass surface density $\Sigma_{g}$ (in a)), the line centroid V$_{o}$ (in b) and the turbulent velocity V$_{turb}$ (in c)), along with their associated 1$\sigma$ uncertainties (in c, d and e). The expected values in a, b and c have been clipped to show only significant values (above 2$\sigma$). \label{Fig:Line}}
\end{figure*}
 
\begin{figure*}[ht!]
\epsscale{0.85}
\plotone{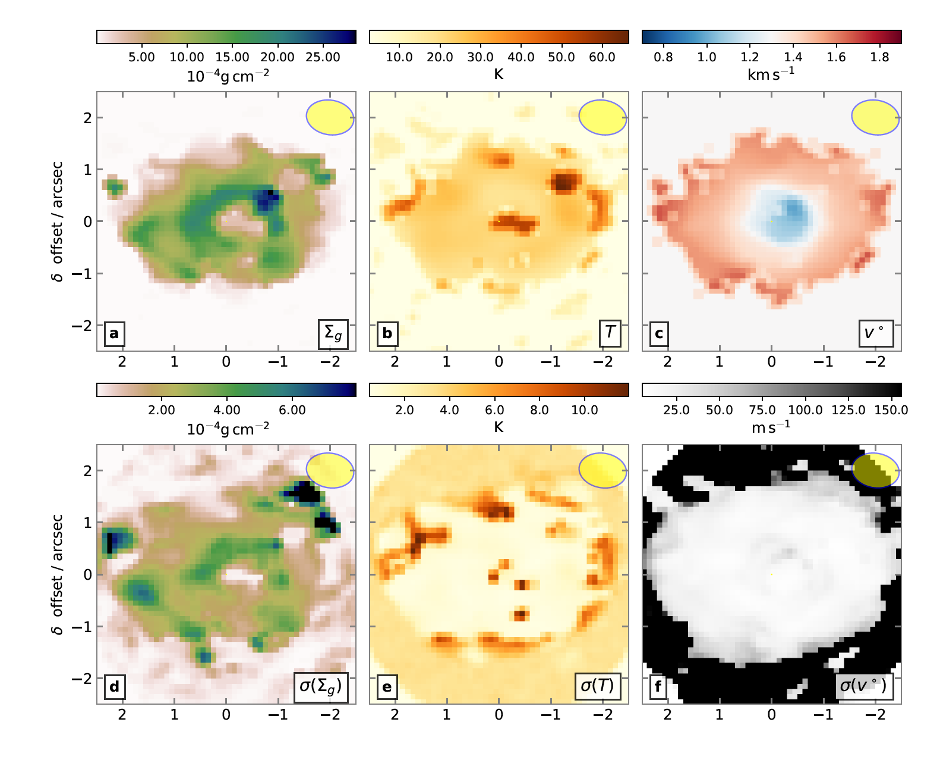}
\caption{Same as Figure \ref{Fig:Line}, but fixing V$_{turb}$ and releasing $\rm T$. \label{Fig:Line2}}
\end{figure*}

In order to test whether the $^{12}$CO ``bubble" is dominated by infall or outflow motions, we proceeded to estimate the CO column densities and total CO mass by using the uniform-slab diagnostics in local thermodynamic equilibrium (LTE). We applied the $\rm SLAB.LINE$ package \citep{Casassus2021} to fit each line of sight of the $^{12}$CO(2-1) and $^{13}$CO(2-1) data cubes by varying the CO column, parameterized in terms of the total molecular mass g (assuming a CO abundance of 10$^{-4}$ relative to H$_{2}$), the LTE temperature $\rm T$, the line of sight velocity turbulence V$_{turb}$ and the line velocity centroid  V$_{o}$. An initial attempt releasing all parameters resulted in strong parameter cross-correlations, where $\rm T$ is degenerate. A full parameter fit requires deeper data, as well as other $^{12}$CO rotational transitions if $^{13}$CO(2-1) remains undetectable. We therefore opted for two complementary approaches: we fixed T = 30 $\rm K$ to obtain an upper limit of the the total CO mass, and then fixed V$_{turb}$ to the channel width to have an idea of the radial profiles of the temperature and the mass. The results of the fit with fixed $\rm T$ are shown in Fig. \ref{Fig:Line}, where the radial distribution of $\Sigma_{g}$ is approximately that of a filled-in shell. The median reduced $\chi^{2}$ where $\Sigma_{g}$ is above 2$\sigma$ is 1.18. The results of the fit by fixing V$_{turb}$ are shown in Fig. \ref{Fig:Line2}, where we see that a central rise in temperature results in more of a shell-like morphology for $\Sigma_{g}$. In this case the median reduced $\chi^{2}$ is 1.35. The total mass obtained with fixed $\rm T$ = 30 K is $\rm M$ = 1.46 $\pm$ 0.01 $\rm M_{\oplus}$ (when summing lines of sight where $\Sigma_{g}$ is detected above 2$\sigma$), with an average density of n($\rm H_{2}$) $\sim$ 10$^{4}$ cm$^{-3}$, assuming that the radius of the shell is $\sim$2 arcsec and with a mean molecular weight = 2.3. Since this estimate uses a fairly low temperature, it represents an upper limit to the total mass of the shell. The corresponding Jeans mass is $\rm M_{Jeans}$ = 2.8 $\times$ 10$^{5}$ $\rm M_{\oplus}$. Thus this shell is not collapsing under its own gravity, and the observed centroid velocity field necessarily corresponds to expansion. 
 
 The $^{13}$CO and $^{12}$CO line fluxes, therefore, indicate column densities of the order of 10$^{-4}$ gr per cm$^{2}$ for the detected ``bubble" -- implying number densities of the order of 10$^{4}$ H$_{2}$ molecules per cm$^{3}$ -- which in turn are 3 orders of magnitude lower than expected for a collapsing core at 100 to 1000 au scales \citep{Aikawa2007}. These densities could only be reconciled with the ALMA observations by assuming extreme CO freeze out, which is inconsistent with the presence of a $>$ 2000 K source at the center of the system (Fig. \ref{Fig:Kmos_spectrum})

\subsection{Scenario 2: The mass loss of a giant star in the distant background (d $>$ 5-10 kpc).} Since the SED and $^{13}$CO/$^{12}$CO line fluxes exclude gravitational collapse in a dense environment, and the distance to SSTc2d J163134.1 is somewhat uncertain, we have also considered the possibility of an expanding shell from an AGB star \citep{Maercker2012, Kerschbaum2017} projected in the far background of the Ophiuchus molecular cloud, i.e. the mass loss of a giant star in the distant background located at $>$ 5-10 kpc. In other words, by inverting the direction of the velocity field and assuming that the central IR source is a giant star located at 10 kpc, the $^{12}$CO data could be interpreted as an expanding shell of $\sim$ 40.000 au across. However, this scenario also has several shortcomings. First, the velocity expansion of the detected shell ($\sim$1 km s$^{-1}$; see Fig. \ref{Fig:ALMA_channel}) is significantly slower than expected for a more common AGB outflowing event. In addition, the size of the ``bubble" would be much larger than typically observed shells associated with AGB stars of the order of 5.000 - 15.000 au across \citep[e.g][]{Loup1993, Winters2003, Cernicharo2014, Maercker2018}.

Furthermore, the probability of SSTc2d J163134.1 being an AGB located around 10 kpc away in the direction of the Ophiuchus Cloud is practically null. First, SSTc2d J163134.1 is best classified as a young dwarf object based on the spectroscopic analysis presented in Sec. \ref{Sec:Spectroscopy}. From a kinematic assessment of membership in Ophiuchus for SSTc2d J163134.1 (see Appedix \ref{App:Proper}), we find that only 1$\%$ of the AGB stars in the Galaxy have similar radial velocities (within 1$\sigma$) as the Ophiuchus members \citep[V$_{\rm HSR}$ = -6.27 $\pm$  1.48 km s$^{-1}$;][]{Esplin2020}, rendering a kinematic chance alignment of an AGB with Ophiuchus even less likely.  If the object belongs to the Ophiuchus cloud at a distance of just $\sim$139 pc, as indicated by its proper motions and radial velocity, its luminosity is well constrained to be $<$ 0.1  L$_{\rm \odot}$. This low luminosity excludes all types of giant stars and implies that SSTc2d J163134.1 must be a very cold dwarf object instead, i.e., a very-low mass star or a brown dwarf.

\section{Most likely scenario: A very-low-mass object with spherical mass loss.}
\label{Sec:Model}

Having ruled out gravitational collapse in a very dense environment and mass loss in a giant star in the far background, we then further explore the novel possibility of a \textit{very-low-mass object with spherical mass loss}. To that end, we used the Line Modeling Engine \citep[LIME;][]{Brinch2010} code to reproduce the $^{12}$CO channel and moment maps along with the PV diagram, see Figs. \ref{Fig:ALMA_channel} and \ref{Fig:Moment}, as further explained below.

\subsection{Expanding Shell Model}
\label{Sec:Expanding}

\begin{figure*}[ht!]
\epsscale{0.85}
\plotone{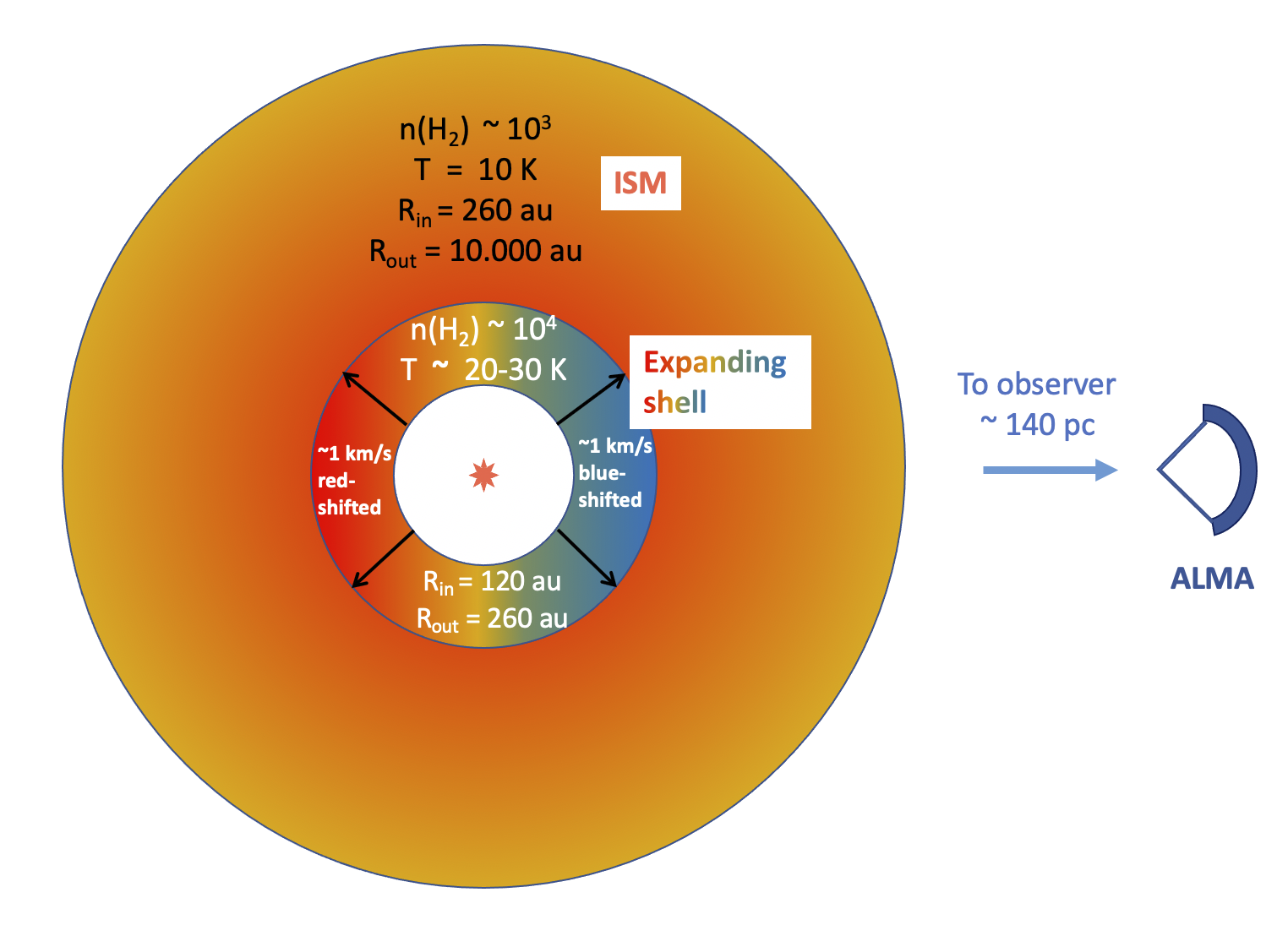}
\caption{Schematic of our  LIME model showing the characteristic sizes,  densities and temperatures of the expanding shell and the surrounding cold material (see Sec. \ref{Sec:Model}).  The blue-shifted portion of the shell and the surrounding cold gas  almost completely absorbs the red-shifted part of the $^{12}$CO emission from the expanding shell. Therefore,  only the blue-shifted part of the shell is detected in the channel maps shown in Fig. \ref{Fig:ALMA_channel}. In the model, the width of the expanding shell is 140 au, a size comparable to the size of the ALMA beam. Therefore, its temperature and density radial profiles are not well constrained by the observations. \label{Fig:Cartoon_model}}
\end{figure*}

To explain the spherical-like shape displayed in the $^{12}$CO channel maps (see Fig. \ref{Fig:ALMA_channel}), we adopt the parameterization given by the  canonical``inside-out" collapse model, i.e., a molecular cloud characterized by an inner region dominated by infalling motions and an outer region in a nearly static phase, which in turn is time-independent \citep{Shu1977}. In order to adjust it to our proposed scenario, i.e. an expanding shell surrounding a central stellar source, we inverted the direction of the radial velocity components while keeping a gas mass density of $\rho$ $\propto$ r$^{-3/2}$ profile . To probe the kinematics of the inner regions of the expanding gas \citep{Terebey1984}, we use the density structure calculated jointly by \citet{Ulrich1976} and \citet{Cassen1981} (hereafter CMU) for a rotationally flattened envelope. The CMU model assumes that the particles fall in from an initial spherical cloud rotating with constant angular velocity, and terminate at the center of the envelope. For our proposed scenario, the particles are expelled with an initial ``maximum" velocity from the center of a spherically symmetric self-gravitating structure. With this assumption, those particles attain lower velocity values outwards from the central source following ballistic orbits due to the gravitational influence of the central mass. The velocity field of an expelled particle along its orbit defined in spherical coordinates ($\rm r$, $\theta$) is given by:

\begin{eqnarray*}
v_r = \left(\frac{GM_{\star }}{r}\right)^{1/2} \left(1+\frac{\cos \theta}{\cos \theta_0}\right)^{1/2},
\end{eqnarray*}

\begin{eqnarray*}
v_{\phi} = \left(\frac{GM_{\star }}{r}\right)^{1/2} \left(1-\frac{\cos \theta}{\cos \theta_0}\right)^{1/2} \left(\frac{\sin \theta_0}{\sin \theta}\right),
\end{eqnarray*}

\begin{eqnarray*}
v_{\theta} = \left(\frac{GM_{\star }}{r}\right)^{1/2} (\cos \theta_0 - \cos \theta) \left(\frac{\cos \theta_0 + \cos \theta}{\cos \theta_0 \sin^2 \theta}\right)^{1/2} ,
\end{eqnarray*}

where $\rm M_{\star}$ is the mass of the central source, $\theta$ is the polar angle of the particle with respect to the rotational axis, and $\theta_{o}$ is the final polar angle of the particle at the end of an orbit. Thus, the radial velocity of a particle along these orbits decreases outwards with a maximum velocity component at a radius of r,

\begin{eqnarray*}
\frac{r}{R_C} = \frac{\sin ^{2}\theta _{o}}{1-\frac{\cos \theta }{\cos \theta_{o}}},
\end{eqnarray*}

where $R_{c}=\frac{R_{o}^{4}\Omega ^{2}}{GM_{\star }}$ is the centrifugal radius, and $R_{max}$ defines the outer radius of the spherical shell. In addition, the density profile for the rotationally-flattened structure within a sphere of radius $\rm r$ is given by:

\begin{eqnarray*}
\rho = \frac{\dot{M}_{\rm out}}{4\pi \sqrt{GM_{\star }R_{c}^{3}}} \left ( \frac{r}{R_{c}} \right )^{-\frac{3}{2}} \left ( 1 + \frac{\cos \theta }{\cos \theta _{o}} \right )^{-\frac{1}{2}} \\
\left ( \frac{\cos \theta }{\cos \theta _{o}}+ \frac{2\cos^{2} \theta _{o}R_{c}}{r}
 \right )^{-1},
\end{eqnarray*}

where $M_{\rm out}$ is the mass outflow rate.

We note the fact that the parameterization of an inverted  ``inside-out" collapse reproduces well the observations of the expanding shell around SSTc2d J163134.1, but does not imply that this is an accurate representation of the physics of the system, which are still unknown. This mostly reflects the fact that \textit{the density, temperature, and velocity radial profiles of the shell are not well resolved or constrained by the  current observations}.  However,  inverting this well-studied model provides a useful approach to obtain a first approximation to the basic properties of the shell: mass, temperature, and width, in addition to the expansion velocity, which can be measured directly from the channel maps.    

\subsection{Building Position-Velocity diagrams.}

To generate data cubes and PV diagram models, we simulate a line of sight that crosses the spherical shell at a position (r, $\theta$, $\phi$), and calculate the velocities that would be observed across the equatorial plane. To that end, the projected velocities for each position along a given line of sight in the y-axis at z = 0 can be expressed as \citep{Tobin2012},

\begin{eqnarray*}
v_{los} = v_{r}(\sin i\sin \Phi_{0} )+v_{\theta }(\cos i\sin \Phi_{0} )+v_{\phi }(\cos \Phi_{0}),
\end{eqnarray*}

where $i$ is the inclination of the rotation axis of the system with respect to the plane of the sky, and $\Phi_{0}$ is the projection angle of the envelope within the line of sight.

Our main goal by generating a set of models is to reproduce the V-shape detected in the position-velocity (PV) diagram (Fig. \ref{Fig:Moment}c), corresponding to half of an expanding spherical shell traced by blue-shifted emission. To that end, it is necessary to place a slab of cold material in front of the line-of-sight causing a strong red-shifted absorption to reproduce the observed ``V-shape" PV diagram. Thus, the absorption signature is determined by an isothermal slab of cold material of T $\sim$ 10 K extending from Rmax ($\sim$260 au) to $\sim$ 10000 au,  with a constant gas mass density  of n(H$_{2}$) of 10$^{3}$ cm$^{-3}$.  The slab of cold gas has radial velocities ranging from 1.3 to 2.3  km s$^{-1}$. The parameters used for the slab of cold material in front of SSTc2d J163134.1 used to reproduce the observations are degenerate, but consistent with the properties of the environment expected in a molecular cloud such as Ophiuchus \citep{Bergin2007}. Figure \ref{Fig:Cartoon_model} illustrates our final adopted model, which combines an expanding shell surrounded by a cloud of intermediate dense material.

\subsection{Fitting the expanding shell model to the $^{12}$CO data}

Following the above approach, the $^{12}$CO line emission was modeled by using the Line Modelling Engine code \citep[LIME][]{Brinch2010}. LIME performs LTE (Local Thermodynamic Equilibrium) line radiative transfer calculations of CO transitions using collisional rate coefficients adopted from \citet{Yang2010}. We adopted a CO abundance of $\rm X$(CO)=10$^{-4}$ relative to H$_{2}$.  Assuming that the system is viewed face-on and is located at 139 pc,  we create a grid by setting a total number of 2000 points as a function of the radial distance from the central object. Thus, we set a computational domain considering the inner ($\rm R_{min}$) and outer ($\rm R_{max}$) radii of the spherical structure. These $\rm R_{min}$ and $\rm R_{max}$ values are the boundaries in which LIME generates a set of random grid points according to the density of the given model (see Sec. \ref{Sec:Expanding}). In the case of $\rm R_{max}$, we opted to use an outer radial extension of 260 au estimated by averaging those radial values obtained along with and perpendicular to the major axis of the $^{12}$CO moment-0 map (Fig. \ref{Fig:Moment}a), while $\rm R_{min}$ is taken as a free parameter in our modelling process, see below. We also assumed an ambient gas density \citep{Bergin2007} of 10$^{3}$ cm$^{-3}$ as a minimum value for the density profile to ensure that no point of the grid falls out of the computational domain. For the gas temperature within the spherical shell, since the $^{12}$CO emission is usually optically thick ($\tau$ $>$1), we make a conservative estimate of the radial temperature profile by determining the maximum temperature of 30 K at the $\rm R_{min}$ location, and assuming a radial power-law dependence,

\begin{eqnarray*}
    \rm T_{gas}=30K \left ( \frac{r}{R_{min}} \right )^{-0.5}.
\end{eqnarray*}

Thus, the model contains 4 adjustable parameters:

\begin{enumerate}
    \item $\rm M_{star}$: Central mass.
    \item $\rm R_{min}$: Inner radial extension of the spherical shell.
    \item $\rm M_{out}$ : Mass outflow rate. 
    \item $\rm R_{c}$: Centrifugal radius.
    
\end{enumerate}

In order to compare the model to our interferometric data, we use the CASA task ``simobserve" to generate synthetic visibilities of the output LIME cube using the uv-sampling of the observed $^{12}$CO emission. The array configuration, integration time, date, and hour angle were set properly to emulate the observing conditions of the real data (see Sec. \ref{Sec:ALMA}). To generate the ALMA-simulated images  from the model visibilities, we used the CASA task ``clean" by applying the same parameters as the observed image. Finally, to examine the molecular line kinematics, we extracted PV diagrams from these images by cutting along a position angle (P.A.) of 45 deg. to ensure a slice covering an average outer radial extension of the observed spherical-like shape emission. This is viable to perform since every extracted PV diagram at different PA values contains the effect of outflowing motion by only differing on their radial extension, i.e. along semi-major axis $\sim$ 310 au and semi-minor axis $\sim$ 200 au. Then, the model and observed PV diagrams are compared to constraint the system parameters able to reproduce the V-emission feature.

\subsection{Obtaining Fitting Parameters.}

\begin{figure*}[ht!]
\epsscale{1.1}
\gridline{\fig{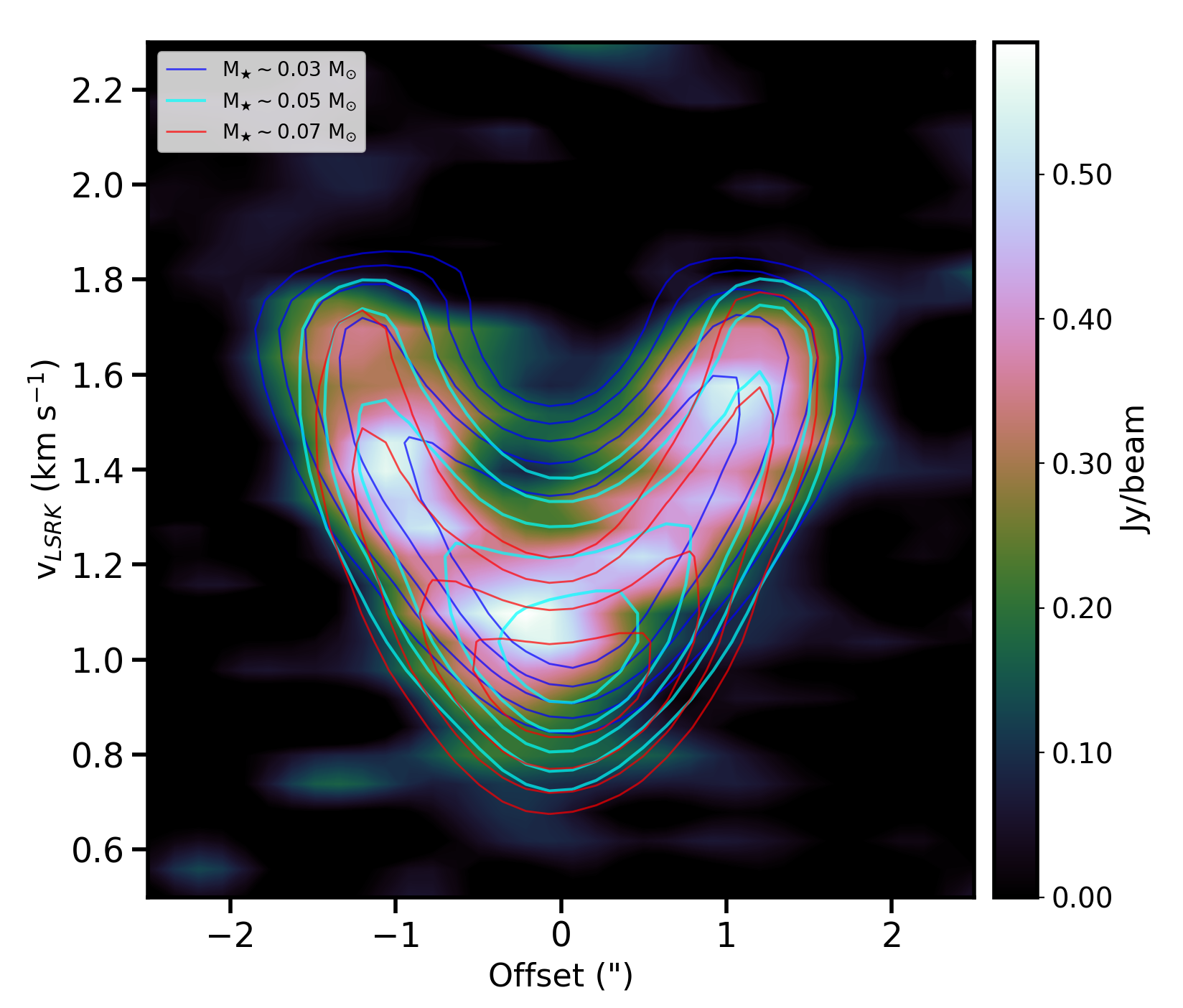}{0.45\textwidth}{(a)}
         \fig{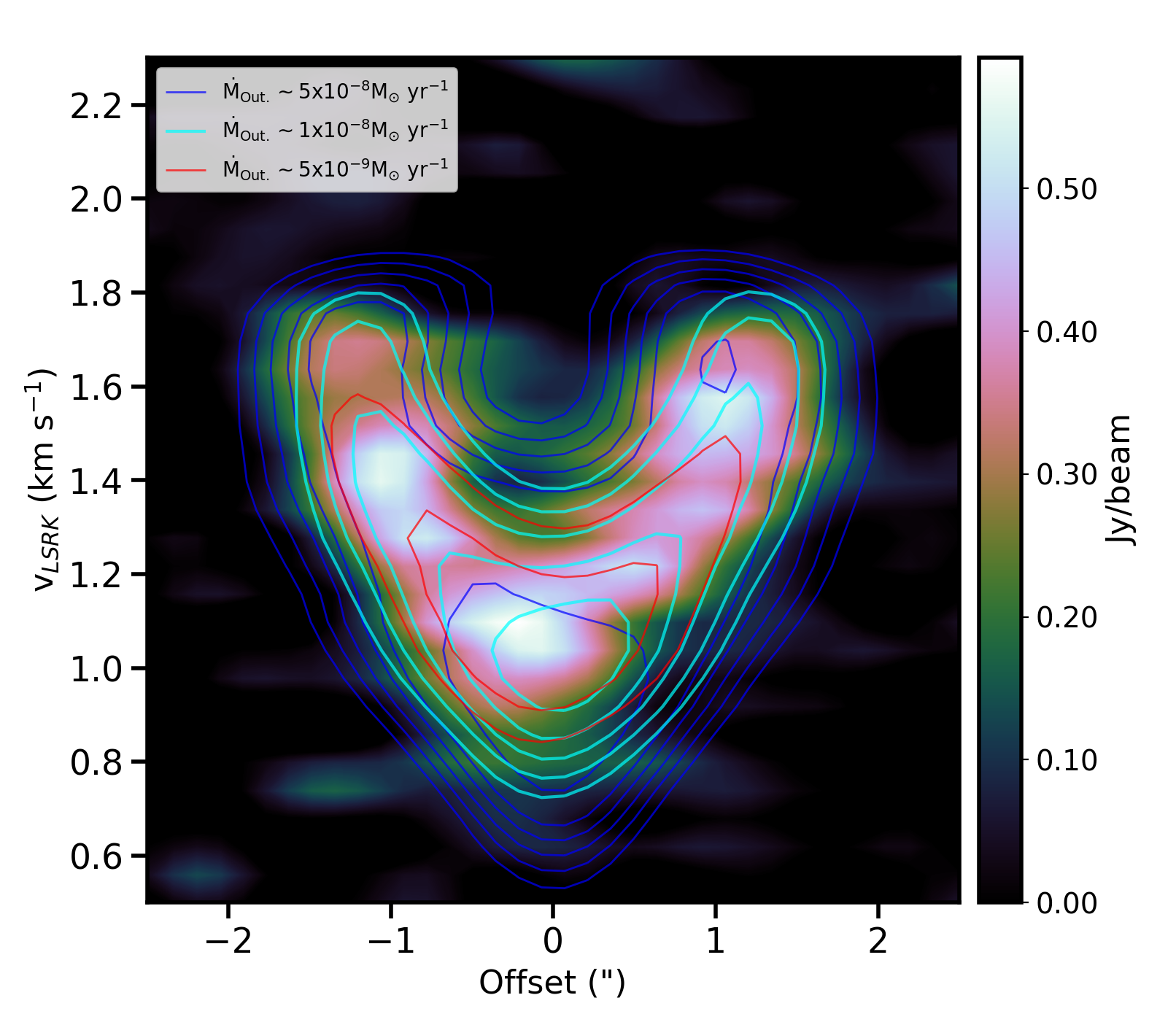}{0.45\textwidth}{(b)}}
\gridline{\fig{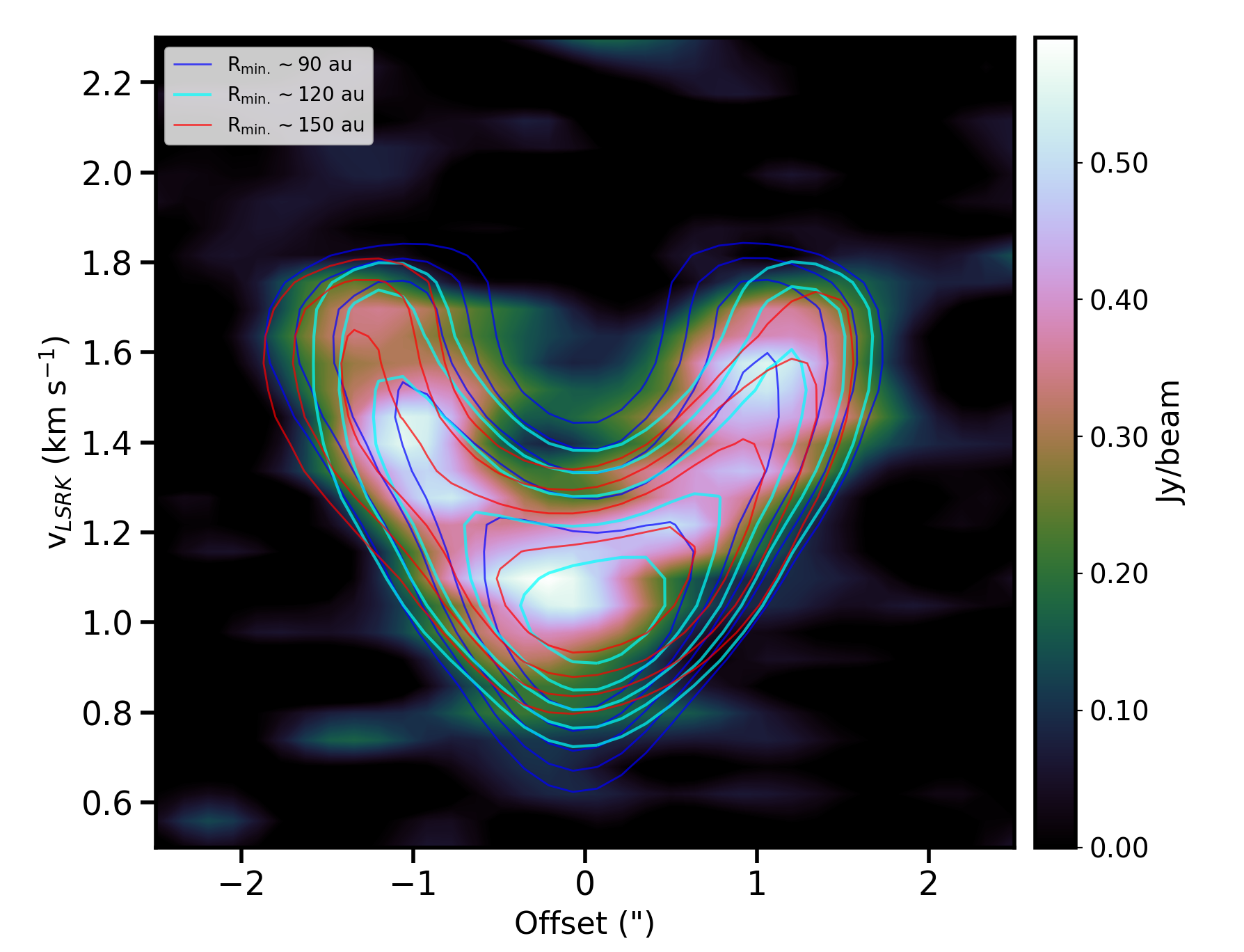}{0.5\textwidth}{(c)}}
\caption{Position-velocity diagrams extracted along a P.A. of 45 deg. for the observed and synthetic (contours) $^{12}$CO emission lines. Overall, the $^{12}$CO emission overlaid with the best model (cyan contours) together with those models reproduced by extreme values corresponding to $\rm M_{\star}$, $\rm M_{out}$, and $\rm R_{min}$. (see Table \ref{Tab:Lime_Best}) and shown in panels a, b, and c, respectively.} \label{Fig:Lime_Best}
\end{figure*}

In our modeling approach, we first explore the impact of $\rm R_{c}$ in the PV diagram model. We find that, in order to generate a detectable V-shape, it is necessary an $\rm R_{c}$ value constrained to $<$ 3 au. For our purposes, $\rm R_{c}$ value was set to 1 au in the radiactive transfer modelling. Notice that the use of a $\rm R_{c}$ $<$ 3 au indicates an insignificant rotation in the shell. The systemic velocity ($\rm V_{sys}$) is taken to be $\sim$ 1.9 km s$^{-1}$ according to the largest extension observed in the $^{12}$CO PV diagram (Figs. \ref{Fig:ALMA_channel} and \ref{Fig:Moment}). We emphasize that $\rm V_{sys}$ does not account for optical effects, and it is likely that we are observing only a section of the spherical-shape,  resulting in an uncertainty in the $\rm V_{sys}$ value. To find a good set of parameters, we started modelling a ring-like shape using a radial extension of $\sim$ 260 au, while iterating over different $\rm M_{out}$, $\rm M_{\star}$, and $\rm R_{min}$ values. For $\rm M_{out}$, we tested values in the range 10$^{-10}$  to 10$^{-3}$ $\rm M_{\odot}$ y$^{-1}$, and varied them as a free parameter in order to scale up or down the intensity of the $^{12}$CO emission. We find that mass-loss rates of larger than 10$^{-8}$ $\rm M_{\odot}$ y$^{-1}$ are required to reach the highest intensity levels, while values lower than 10$^{-7}$ $\rm M_{\odot}$ y$^{-1}$ are needed reproduce the desired V-shape (see Fig. \ref{Fig:Lime_Best}). In the adopted model, the central mass controls the the velocity field, and we explore a broad range of values between 0.01 $\rm M_{\odot}$ and 2.0  $\rm M_{\odot}$, while for $\rm R_{min}$. , we explore values between 10 au and 250 au. Overall, $\rm M_{\star}$ and $\rm R_{min}$ were varied to better match the data and adjust the V-shape in the PV diagram. After comparing every model in our grid of models to the observed emission, we find that $\rm M_{\star}$ and $\rm R_{min}$. significantly impact the velocity gradient, thus placing a limit on those values. In the case of $\rm M_{\star}$, values $>$ 0.05 $\rm M_{\odot}$ displace the $^{12}$CO emission towards larger red- and blue-shifted values and from interior region outwards, altering the V-shape. In addition, a central mass of $\rm M_{\star}$ $>$ 0.05 $\rm M_{\odot}$ reduces intensity levels in the model, whose value has been challenging to increase to the observed levels. Thus, the total central mass cannot be substantially larger than 0.08 $\rm M_{\odot}$ without being inconsistent with the $^{12}$CO data or the model. Similarly, $\rm R_{min}$ values smaller than 90 au or larger 150 au (i.e., at increasing  or reducing the thickness of the expanding shell) distort the V-shape of the PV diagram by closing ($<$ 90 au) or opening ( $>$150 au) the internal region of the emission (Fig. \ref{Fig:Lime_Best}). The best-fit values obtained as described above and the range of values explored in our grid of models are summarized in Table \ref{Tab:Lime_Best}.

\begin{figure*}[ht!]
\epsscale{1.0}
\plotone{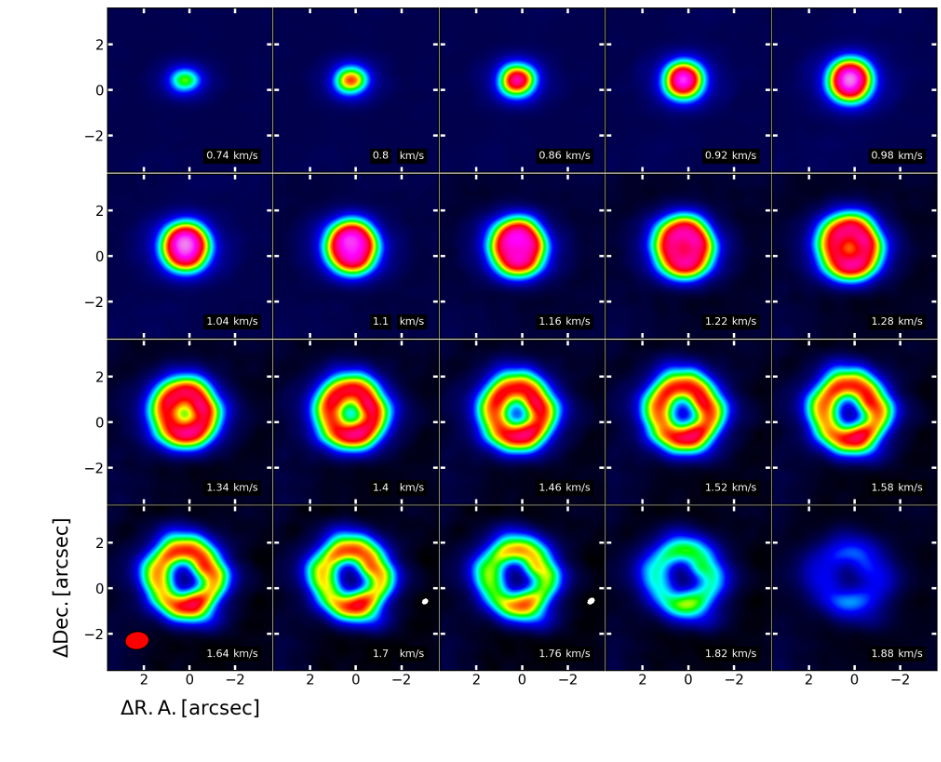}
\caption{Velocity channel maps of the best synthetic $^{12}$CO emission line towards SSTc2d J163134.1. The velocity of the channels is indicated in the lower right corner of each panel, which are the same range as in Fig. \ref{Fig:ALMA_channel}.} \label{Fig:Channel_Best}
\end{figure*}

\begin{figure*}[ht!]
\epsscale{1.0}
\gridline{\fig{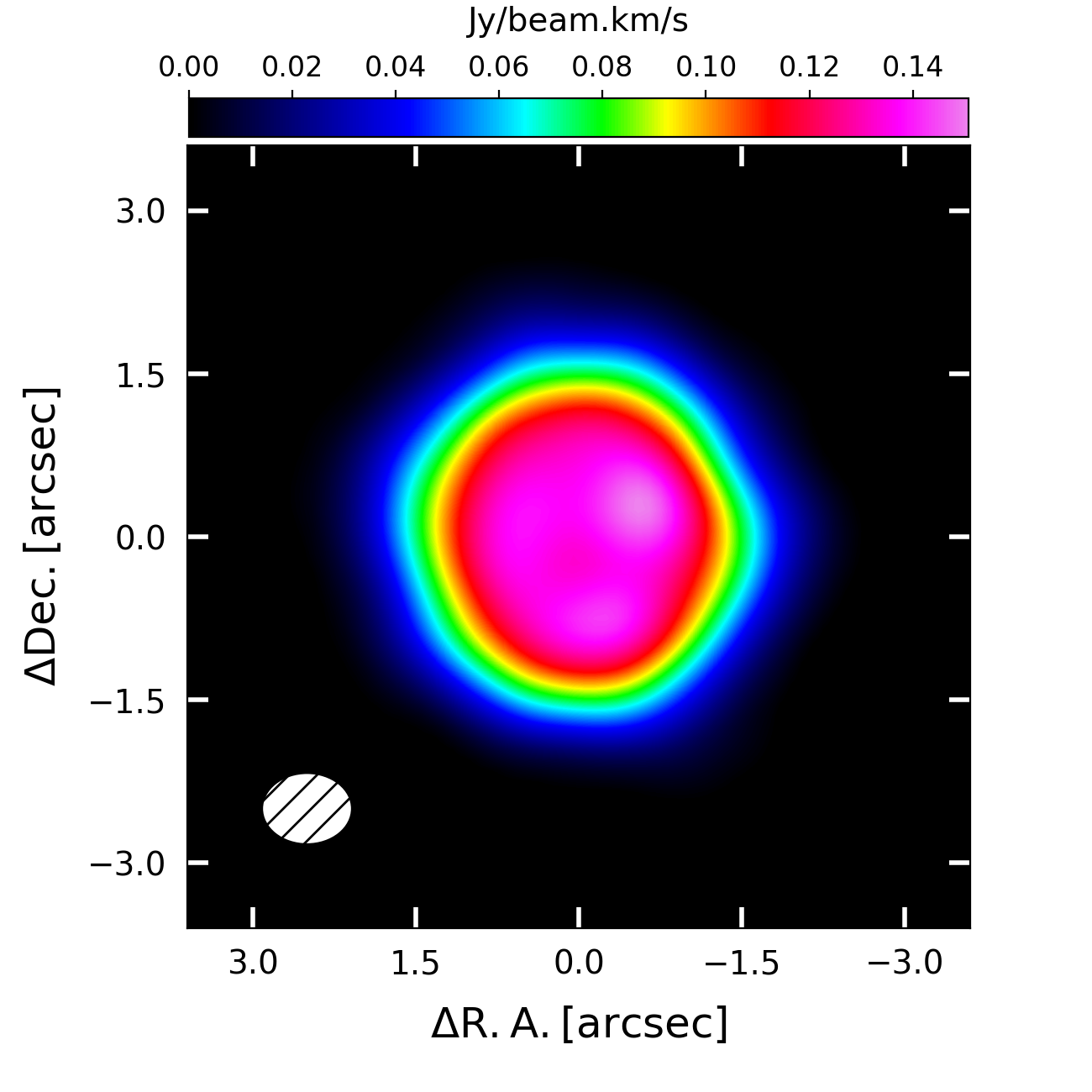}{0.5\textwidth}{(a)}
         \fig{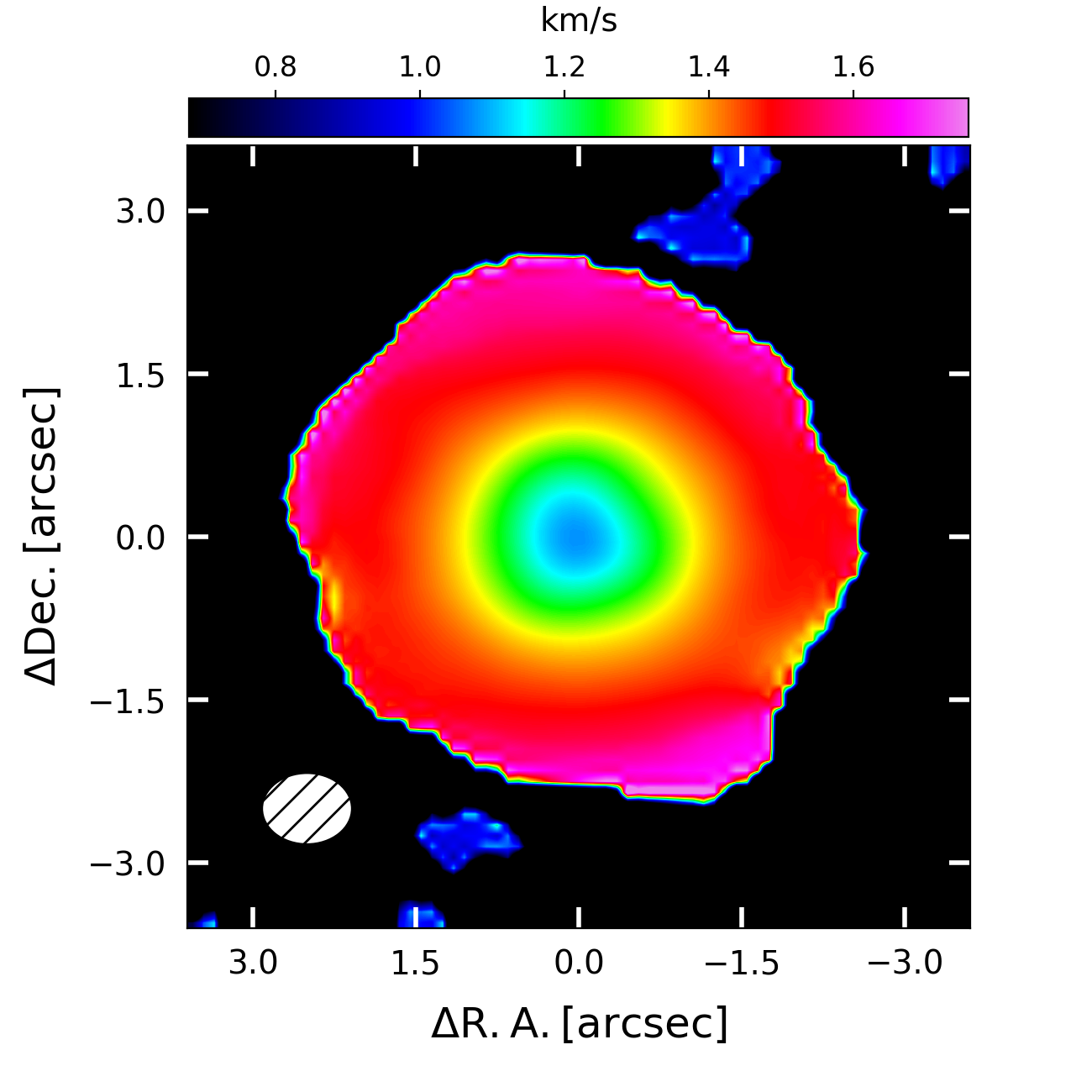}{0.5\textwidth}{(b)}}
\caption{Synthetic $^{12}$CO integrated intensity (a) and intensity weighted mean velocity (b) maps. The integrated emission of the synthetic line was performed in the range from 0.7 to 1.88 km s$^{-1}$. In each panel, the corresponding beam size is shown at the bottom left.} \label{Fig:Moment_Best}
\end{figure*}

The PV diagram of the best-fit models is shown in Fig. \ref{Fig:Lime_Best}. The synthetic velocity channel and moment maps of the best synthetic model are also shown in Figs. \ref{Fig:Channel_Best} and \ref{Fig:Moment_Best}, respectively. Overall, we find that our simple model of an expanding shell embedded in a cold ISM reproduces well the observations, supporting the proposed physical phenomenon:  \textit{a shell of gas expelled from a young brown dwarf}.

\begin{deluxetable*}{cccc}
\tablenum{1}
\tablecaption{Best-fit parameters for parametric models fit to ALMA $^{12}$CO(2–1) data, which match those predicted for outflowing motion, see Fig. \ref{Fig:Lime_Best} \label{Tab:Lime_Best}}
\tablewidth{0pt}
\tablehead{
\colhead{Parameter} & \colhead{Best Fit} & \colhead{Min-Max Limits} & \colhead{Range of explored values} }
\startdata
$M_{\star}$ [$M_{\odot}$] & 0.05 & 0.03 to 0.07 & 0.01 to 2.0  \\
$\rm R_{min}$ [au] & 120 & 90  to 150 & 1 to 250 \\
$M_{out}$ [$M_{\odot}$ yr $^{-1}$]  & 1 $\times$ $^{-8}$  &  5 $\times$ $^{-8}$ to 1 $\times$ $^{-9}$  & 1 $\times$ $^{-6}$ to 1 $\times$ $^{-10}$\\
\enddata
\end{deluxetable*}

\section{Discussion} \label{Sec:Discussion}

\subsection{A deuterium-flash Brown Dwarf? }

Strong quasi-spherical mass loss as seen in SSTc2d J163134.1 is not known to occur in young brown dwarfs. However, the structure and evolution of young BDs remain uncertain \citep{Burrows2001, Luhman2012}, and by analogy to the thermal pulses that drive the mass loss in AGB stars \citep{Schwarzschild1965}, it can be hypothesized that the onset of deuterium burning \citep{Palla2005, Bodenheimer2013} in a young BD could result in a thermal pulse strong enough to eject part of its atmosphere.  Potential instabilities induced by deuterium-burning in very low mass stars and brown dwarfs have been proposed over 50 years ago \citep{Gabriel1964} but they still lack observational support \citep{Cody2014}.  Deuterium burning requires core temperatures of around 3$\times$10$^{5}$ K \citep{Bodenheimer2013}, and thus the age at which the burning starts depends on the mass of the object. On the one hand, for objects with masses $>$ 0.1 $\rm M_{\odot}$, the onset of deuterium burning is expected to occur at a very young age ($<$ 0.5 Myr) while the objects are still strongly accreting and deeply embedded in their circumstellar envelopes; i.e. Class-I phase \citep{Palla2005, Evans2009}  making it difficult to identify. On the other hand, for masses between 20 and 80 $\rm M_{Jup}$, deuterium burning should occur at ages between $\sim$1 and 10 Myr \citep{Palla2005}, which encompasses the characteristic age range of (sub-)stellar objects in the Ophiuchus molecular clouds ($\sim$1-3 Myrs), and other star-forming regions \citep{Cieza2007}, rendering the phenomenon more likely to be detected at later evolutionary phases in low-mass stars and brown dwarfs.

Just before the onset of deuterium burning (at an age of $\sim$0.6 Myr), a 80  $\rm M_{Jup}$  brown dwarf should have an effective temperature of $\sim$2850 K, and a luminosity of $\sim$0.07 $\rm L_{\odot}$.  Similarly, a 20 $\rm M_{Jup}$ brown dwarf should have a 2570 K effective temperature and a 0.003 $\rm L_{\odot}$ luminosity when deuterium burning starts at an age of around 2 Myr \citep{Baraffe2015}. In the case of SSTc2d J163134.1, we estimate a luminosity of $\sim$0.19 $\rm L_{\odot}$ assuming a K-band bolometric correction for young M dwarfs \citep{Filippazzo2015}, a distance of $\sim$ 139 pc, A$\rm _{v}$ = 42.2, and a temperature of $\sim$2770 K based on its spectral type \citep{Herczeg2014}. Assuming a membership to L1709 with an age of $\sim$ 2 Myrs \citep{Esplin2020}, SSTc2d J163134.1 is considerably brighter than the predictions of standard evolutionary models \citep{Baraffe2015}. For instance, for a mass of 50 $\rm M_{Jup}$ at an age of 2 Myr, the \citet{Baraffe2015} models predict a temperature of $\sim$2830 K, a radius of $\sim$0.6 $\rm R_{\odot}$, and a luminosity of $\sim$ 0.02 $\rm L_{\odot}$. Although, the mass of SSTc2d J163134.1 is not well constrained yet (see Sec. \ref{Sec:Model}), if SSTc2d J163134.1 is in fact a 50 $\rm M_{Jup}$ brown dwarf  with an age of $\sim$2 Myr, this would imply an increase in radius by a factor of three with respect to standard model predictions to explain its relatively high luminosity for the given effective temperature.

Model predictions for young brown dwarfs should be taken with a lot of caution because deuterium burning is not yet well understood, as it is highly dependent on the mass of the object and even its formation history \citep{Molliere2012}. The energy generation rate, in particular, is much more difficult to estimate than the age at the onset of deuterium burning because it scales with the 12-th power of the temperature \citep{Stahler1988} and is therefore extremely sensitive to the thermal structure of the object and the region in which deuterium burning can occur.  Some detailed models of deuterium burning in giant planets and low-mass brown dwarf  ( $\rm M$ $\sim$ 10–30 $\rm M_{Jup}$) do find an expansion of the photospheres during the deuterium-burning phase and an increase in luminosity by one to three orders of magnitude \citep{Molliere2012, Bodenheimer2013}, in general agreement with our observational constraints.

In addition to the constraints on temperature and luminosity evolution provided by SSTc2d J163134.1, the ejected shell of gas around it suggests a short and violent event analogous to the helium shell flashes that power the thermal pulses and mass loss in AGB stars \citep{Mattsson2007}. SSTc2d J163134.1 could thus be considered the prototype of a ``deuterium-flash brown dwarf" \citep{Salpeter1992}, a new class of brown dwarf observationally characterized by a high luminosity and (quasi-) spherical mass-loss. Accordingly, overluminosity should be the most conspicuous feature of this class of brown dwarf; however, while some young over luminous brown dwarfs exist in the literature \citep[e.g. USco 1610-1913 B, USco 1602-2401 B, 2MASS J16224385-1951057; ][]{Petrus2020, Aller2013, Dahm2012}, none of them are as over-luminous as SSTc2d J163134.1 or show spherical mass loss. Comparing SSTc2d J163134.1 to other late M- and L- type members of the Ophiuchus cloud complex, SSTc2d J163134.1 belongs to the group of the brightest members in this star-forming region, see Appendix \ref{App:HR}. From this small sample of over-luminous objects (see Table \ref{Table:HR}), whose spectral range covers between late M0 and M8 types, SSTc2d J163134.1 stands out as a bright and disk-less object located in the L1709 region while only a couple of objects located in L1688 (2MASS J16272183-2443356, 2MASS J16262523-2423239) exhibit similar observational features as our target. Unfortunately, none of these objects have been observed at (sub-)millimeter wavelengths to confirm or rule out the presence of spherical mass-loss.

The fact that SSTc2d J163134.1 is the first brown dwarf known to exhibit both overluminosity and mass-loss  suggests that  the ``deuterium-flash" phase is very short and/or that not all brown dwarfs go through this phase. This is supported by the very short dissipation timescales of the $^{12}$CO ``bubble" shape of about 10$^{4}$-10$^{5}$ yrs, assuming overall spherical symmetry and steady flows. Furthermore, from the size of the  known population of brown dwarfs in nearby (d $<$ 300 pc) star-forming regions ($\sim$10$^{3}$) and their characteristic ages (a few $\times$ 10$^{6}$ yr), the duration of the deuterium-flash phase can be constrained to be only 10$^{4}$-10$^{5}$ yrs. Given these constraints, as well as giving conditions such as over-luminosity in disk-less objects, the identification of additional deuterium-flash brown dwarfs might require deep near-IR and mm observations in more distant regions, such as the Orion Nebula Cluster, containing very large populations of substellar objects. In the future, the detailed study and characterization of SSTc2d J163134.1 and other deuterium-flash brown dwarfs will provide much-needed observational guidance on the internal structure and luminosity evolution of young brown dwarfs.  

\subsection{A companion being engulfed by a brown dwarf?}

Although the deuterium-flash is the proposed explanation for the observed mass loss,  such scenario still needs to be explored by theoretical work.  More exotic scenarios  can not be ruled out at this point. For instance, another possibility that might explain an expelling shell due to a sudden increase of the mass-loss rate could be tidal interactions between the host BD and a planetary companion. It has been suggested that planets that closely interact with the host star may get deviated from their orbits causing the planet engulfment onto the star at some time during the evolution of the system. Such accretion process of a planet onto the core is accompanied by a substantial expansion of the star that can lead to a high mass ejection \citep{Siess1999a, Siess1999b}. The rate of this ejection of stellar material depends on the star’s and planet’s mass and size, as well as the orbital eccentricity. Once the planet is finally consumed by the star, it would be expected large ejections of material, either resulting in winds and/or non-spherical, dipole-shaped planetary nebulae \citep{Stephan2020}. However, the planet engulfment scenario has not yet been explored with young brown dwarfs as central objects. While the observation of an active consumption event might be challenging, other signatures such as strong stellar wind, as well as strong optical, UV, and X-ray radiation can be expected if a planet entering the star’s atmosphere is generating the observed ``bubble" shape emission. Future X-ray and UV observations may detect some of the predicted features of a planet being consumed by its host star and thus, able to explain the presence of a ``bubble" CO emission shape located in the Ophiuchus star-forming region.

\section{Summary and Conclusions}
\label{Sec:Summary}

We have reported the serendipitous discovery of an expanding shell of carbon monoxide (CO) ejected from a $<$ 3000 K object located in the direction
of the Ophiuchus Molecular Cloud. This unique system was observed as part of the ``Ophiuchus Disk Survey Employing ALMA" (ODISEA) program at a resolution of $\sim$ 0.8$^{"}$. From these observations, a bright ellipse of $^{12}$CO emission of $\sim$ 3$^{"}$ $\times$ 4$^{"}$ is detected towards SSTc2d J163134.1, while ALMA fails to detect 1.33 mm continuum, $^{13}$CO and C$^{18}$O emission. We have shown that the elongated or $^{12}$CO ``bubble"-like emission shape is actually associated with SSTc2d J163134.1, a young M7 star, which has been classified using IR spectra taken with the VLT-KMOS instrument. Based on the proper motions of the Ophiuchus members and SSTc2d J163134.1, it is highly likely that this object is located at a distance of $\sim$139 pc. In order to explain the nature of SSTc2d J163134.1, we have explored two different scenarios: ``the inside-out collapse of a dense molecular core in the Ophiuchus cloud" and ``the mass loss of a giant star in the distant background (d $>$ 5-10 kpc)". However, these known astrophysical phenomena present several shortcomings and fail to fulfill the physical characteristics to describe the observations. For that matter, we have proposed an alternative physical phenomenon, \textit{a shell of gas expelled from a young brown dwarf}, which is located right behind a very dense region of the Ophiuchus Molecular Cloud. We suggest that a deuterium flash could be responsible for the observed phenomenon, but detailed theoretical work is needed to demonstrate the feasibility of such a scenario. Other exotic scenarios such as the  engulfment of planets can not be ruled out and are worth exploring in the future.

\begin{acknowledgments}
This paper makes use of the following ALMA data: ADS/JAO.ALMA2016.1.00545.S. ALMA is a partnership of ESO (representing its member states), NSF (USA) and NINS (Japan), together with NRC (Canada), MOST and ASIAA (Taiwan), and KASI (Republic of Korea), in cooperation with the Republic of Chile. The Joint ALMA Observatory is operated by ESO, AUI/NRAO and NAOJ. L.A.C acknowledges support from FONDECYT REGULAR 1211656, Agencia Nacional de Investigación y Desarrollo de Chile. S.C.  acknowledges support from Agencia Nacional de   Investigaci\'on y Desarrollo de Chile (ANID) given by FONDECYT Regular grants 1211496  and ANID project Data Observatory Foundation  DO210001.
\end{acknowledgments}

\bibliography{sample631}{}
\bibliographystyle{aasjournal}

\clearpage

\appendix

\section{Spectral classification}
\label{App:Spectra}

\begin{figure*}
\epsscale{0.7}
\plotone{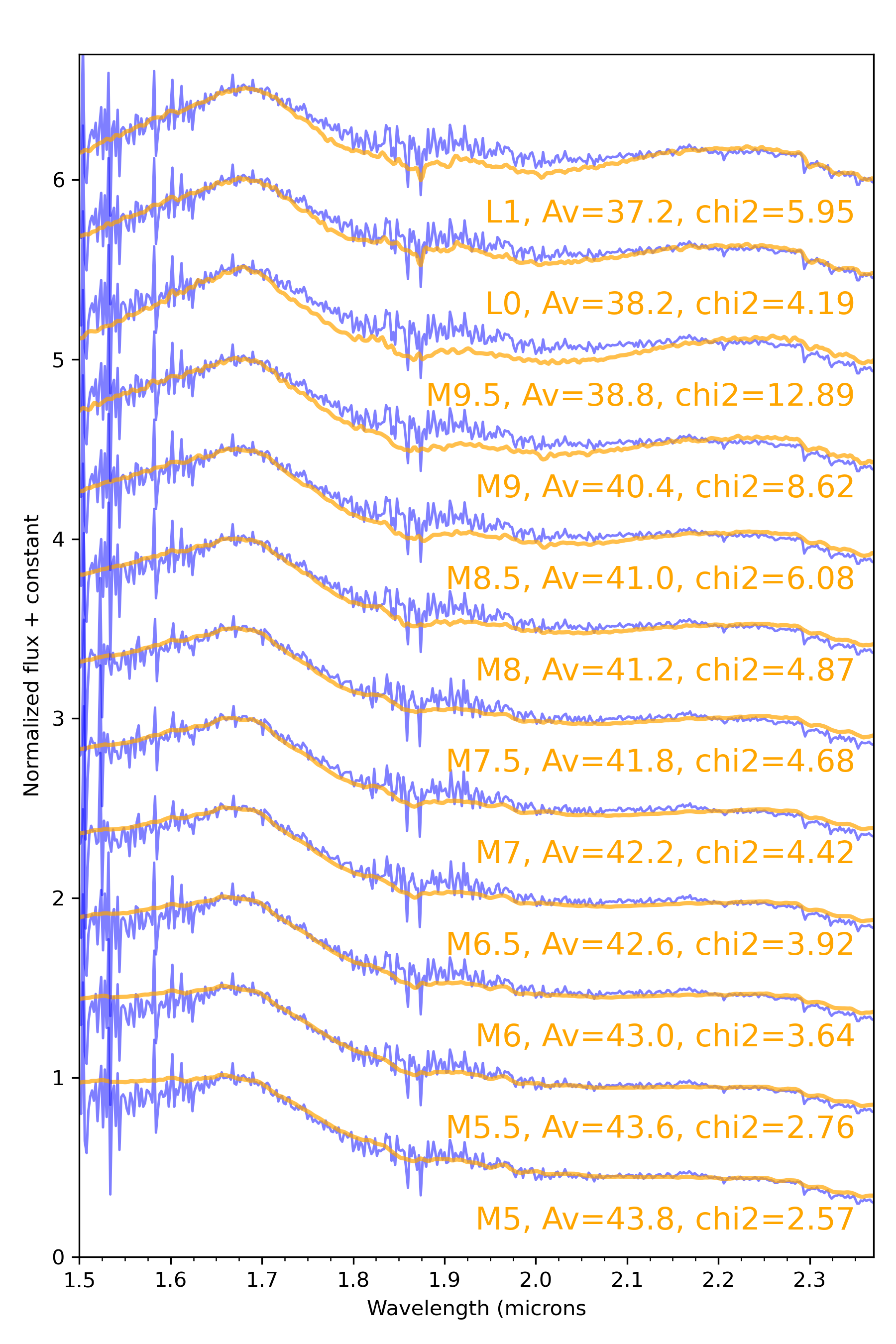}
\caption{Sample of the best template spectra for young M and L spectral types used in our spectroscopic analysis (see Sec \ref{Sec:Spectroscopy}), and compared to the KMOS spectrum of SSTc2d J163134.1 (violet color). The spectra have been normalized at 1.66 $\micron$, and a constant is added to each template to improve readability. The spectrum of SSTc2d J163134.1 has been de-reddend by the corresponding extinction.
\label{Fig:Spectra_young}}
\end{figure*}

\begin{figure*}
\epsscale{0.7}
\plotone{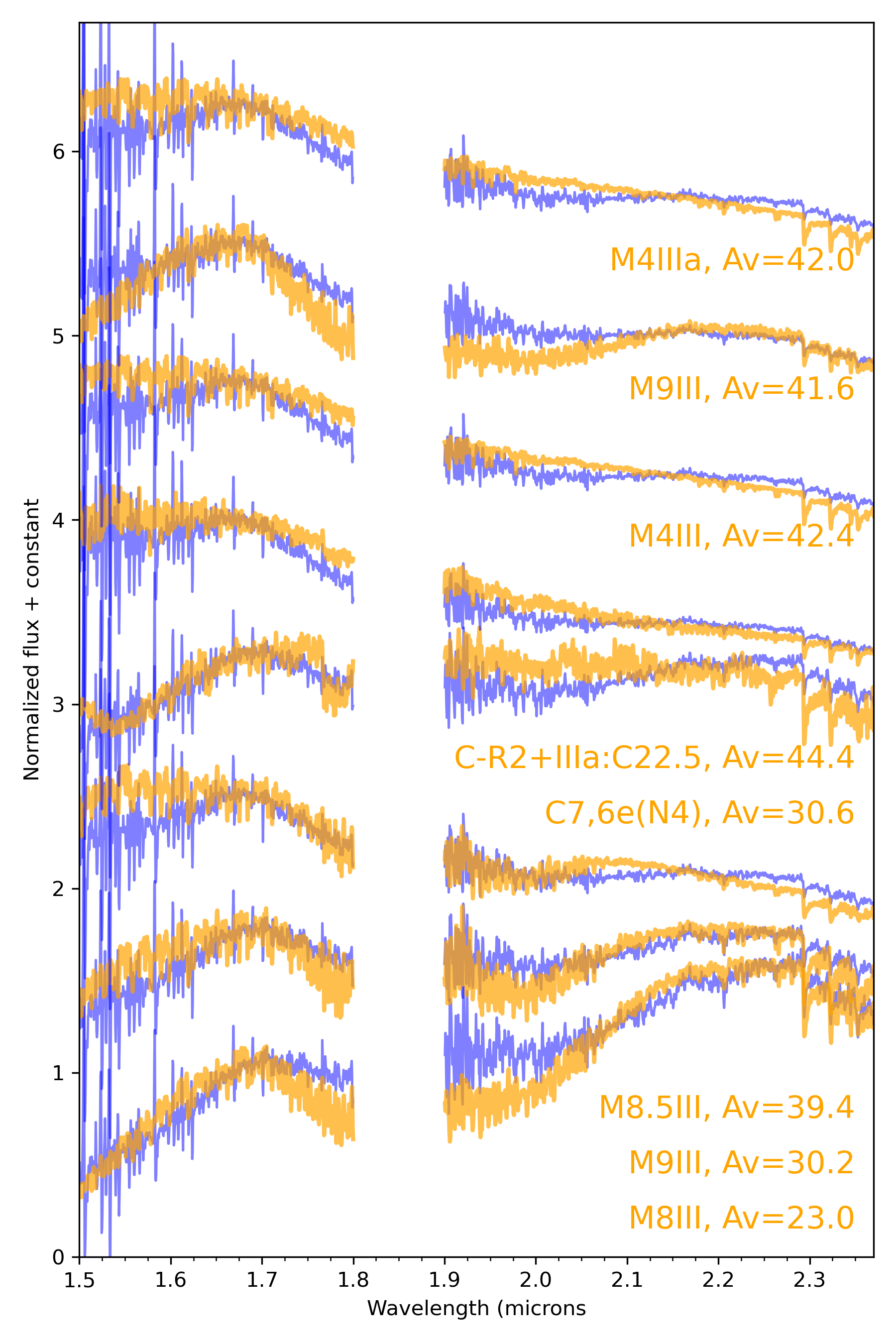}
\caption{Sample of the best template spectra for giant spectral types used in our spectroscopic analysis (see Sec \ref{Sec:Spectroscopy}), and compared to the KMOS spectrum of SSTc2d J163134.1 (violet color). The spectra have been normalized at 1.66 $\micron$, and a constant is added to each template to improve readability. The spectrum of SSTc2d J163134.1 has been de-reddend by the corresponding extinction. \label{Fig:Spectra_giants}}
\end{figure*}

\begin{figure*}
\epsscale{0.7}
\plotone{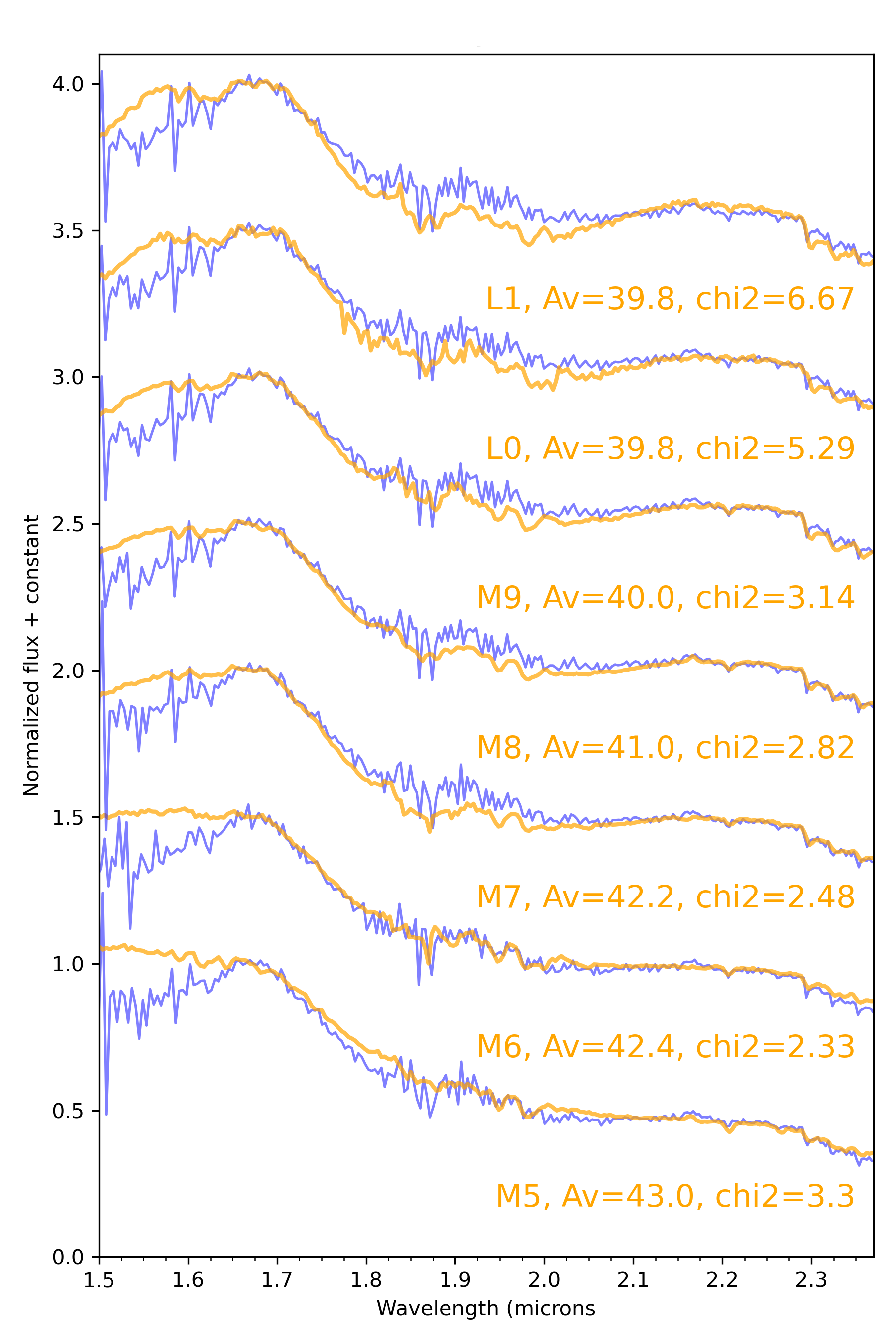}
\caption{Sample of the best template spectra for field M and L spectral types used in our spectroscopic analysis (see Sec \ref{Sec:Spectroscopy}), and compared to the KMOS spectrum of SSTc2d J163134.1 (violet color). The spectra have been normalized at 1.66 $\micron$, and a constant is added to each template to improve readability. The spectrum of SSTc2d J163134.1 has been de-reddend by the corresponding extinction. \label{Fig:Spectra_field}}
\end{figure*}

The classification of M- and L-type stars is dependent on the identification of particular molecular features in the stellar spectrum. The bandheads of the main molecular features and other lines used for the classification of our KMOS spectrum presented in Sec. \ref{Sec:KMOS} are described in Sec. \ref{Sec:Spectroscopy}. Figures \ref{Fig:Spectra_young}, \ref{Fig:Spectra_giants}, and \ref{Fig:Spectra_field} display the best spectral match found through direct comparison of SSTc2d J163134.1 to the young, field and giant standard spectra from \citet{Luhman2017, Kirkpatrick2010, Cushing2005}, and \citet{Rayner2009}. The KMOS spectrum of SSTc2d J163134.1 has been dereddened to match the slope of the selected template library.

\section{A kinematic assessment of membership in Ophiuchus for SSTc2d J163134.1}
\label{App:Proper}

\begin{figure*}
\epsscale{0.7}
\plotone{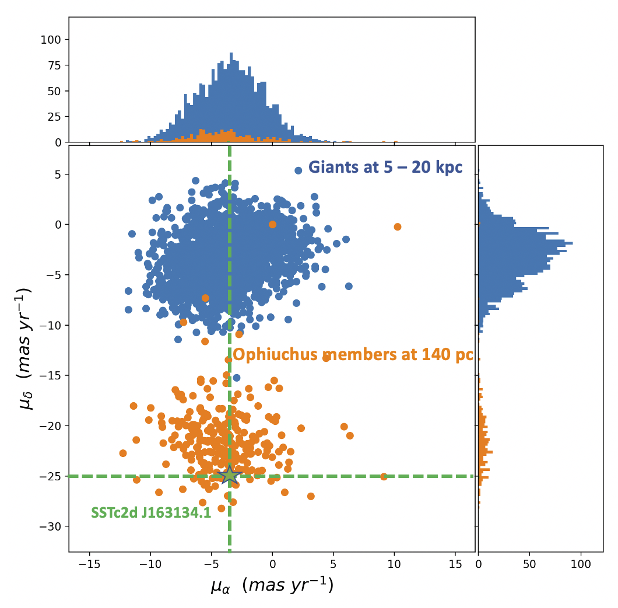}
\caption{Proper motions for Ophiuchus members (orange circles) at 139 pc obtained using Spitzer-IRAC data \citep{Esplin2020}. The proper motions of SSTc2d J163134.1 (green cross hairs) match those of Ophiuchus members and are highly inconsistent (at the 8$\sigma$ level) with the proper motions expected for background objects at 5-20 kpc (blue circles) based on the Besancon Galaxy model \citep{Robin2003}. \label{Fig:Mahalonis}}
\end{figure*}

In order to further constrain or rule out the probability of the star being an AGB located around 10 kpc away in the direction of the Ophiuchus Cloud, we ran a dynamical simulation of stellar populations using the Besancon Galaxy model \citep{Robin2003}. The simulation included stars located between 5 and 20 Kpc (based on the extinction-corrected K-band magnitude of 10.1 of SSTc2d J163134.1)  in a solid angle of 10 degrees square around the coordinates of SSTc2d J163134.1 (l = 354.17402 deg, b = $+$16.27924 deg). The extinction model chosen was the \citet{Marshall2006} model without discrete clouds. Only cool (Teff $<$ 4700 K) giant stars with radii larger than 10 $\rm R_{\odot}$ were considered. The Besancon Galaxy model generates positions, velocities, colors, magnitudes, and stellar parameters of the stars. We find that none of the over 2000 giant stars in the model have proper motions as high as SSTc2d J163134.1, see Fig. \ref{Fig:Mahalonis}. The resulting kinematic center of the giant stars in the model is a -3.47 mas yr$^{-1}$ in RA and a -3.18 mas $^{-1}$ in Dec, with a 1$\sigma$ dispersion of 4.12 mas $^{-1}$ in RA and 3.83 mas $^{-1}$ in Dec.  Following the Mahalanobis distance approach \citep{Mahalanobis1936} as a tool for identifying multivariate outliers, it is found that SSTc2d J163134.1 is located at 8$\sigma$ from the kinematic center of the giants sample. Then, the probability of finding a giant star above the Galactic Center at a distance of 5 to 20 kpc and with the same proper motions as SSTc2d J163134.1 is extremely small ($<$ 8$\sigma$). This means that the proper motions \citep[RA$_{\rm pm}$ = -3.60 $\pm$ 2.04 mas yr$^{-1}$, Dec$_{\rm pm}$ = -25.03 $\pm$ 2.41 mas yr$^{-1}$;][]{Esplin2020} and radial velocity (V$_{\rm LSR}$ = 2.0 km s$^{-1}$, V$_{\rm HSR}$ = -7.38 km s$^{-1}$) of SSTc2d J163134.1 match those of the Ophiuchus members \citep{Prato2007, Canovas2019, Esplin2020}, confirming membership to the cloud.

\section{HR Diagram: Evolutionary Analysis}
\label{App:HR}

\begin{figure*}
\epsscale{0.9}
\plotone{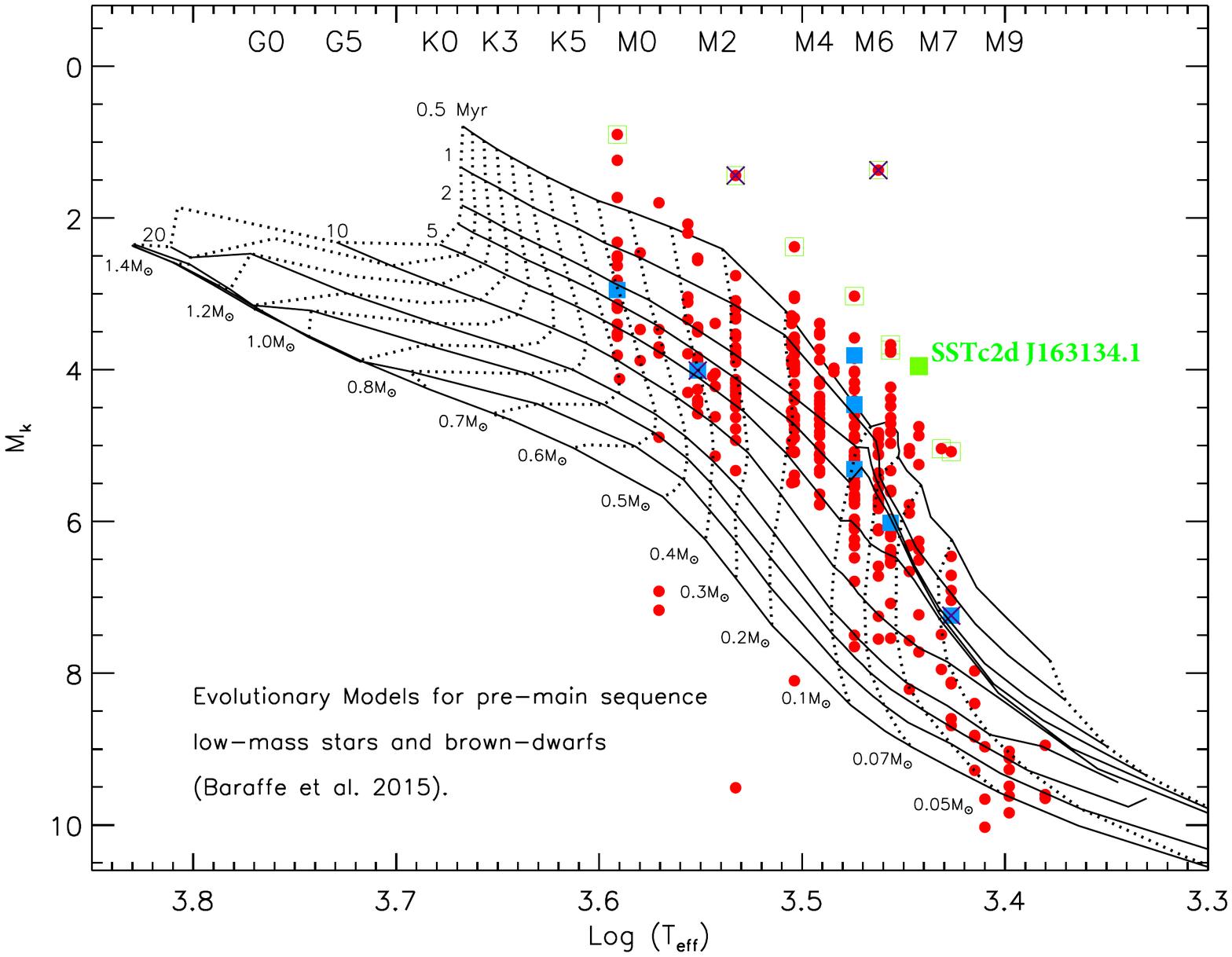}
\caption{T$\rm _{eff.}$– M$\rm _{k}$ diagram of all young M and L dwarf members of the Ophiuchus cloud complex (Red circles) identified by \citet{Esplin2020}. The Ophiuchus members are compared to the BT-Settl isochrone and track models calculated by \citet{Baraffe2015}. Blue squares represent the L1709 members while green open squares represent the brightest objects in the Ophiuchus star-forming region based on their M$\rm _{k}$ at a distance of $\sim$139 pc. The green square shows the position of SSTc2d J163134.1 in the HR diagram, where is evidently its overluminosity compared to other members in the region. Black crosses show those objects that are described as ``nodisk" due to their lack of IR excess, see Table \ref{Table:HR}. \label{Fig:HR}}
\end{figure*}

Figure \ref{Fig:HR} displays the SSTc2d J163134.1 system and the young M and L dwarf members of the Ophiuchus cloud complex identified by \citet{Esplin2020} compared to the BT-Settl isochrone and track models calculated by \citet{Baraffe2015}. Using absolute K-band magnitudes (M$_{K}$) at an average distance of $\sim$139 pc, and spectral types converted to T$\rm _{eff}$ from the conversion scale of \citet{Herczeg2014}, we find that SSTc2d J163134.1 is the brightest M7 type low-mass star from the L1688 (2 Myrs), L1689 (2 Myrs), and L1709 (2 Myrs) members. Table \ref{Table:HR} shows the stellar parameters and K-band photometry of the L1709 and brightest young M dwarf members (L1688) of the Ophiuchus cloud complex \citep[][and references therein]{Esplin2020}.

\begin{deluxetable*}{cccccccc}
\tablenum{1}
\tablecaption{L1709 and brightest young M dwarf members (L1688) of the Ophiuchus cloud complex identified by \citet{Esplin2020}. \label{Table:HR}}
\tablewidth{0pt}
\tablehead{
\colhead{2MASS} & \colhead{R.A.} & \colhead{Dec.} & \colhead{SpT} & \colhead{K-band} & \colhead{A$\rm _{k}$} &\colhead{Disk Type$^{a}$} & \colhead{Region} \\
 & [deg.] & [deg.] &  & [mag] & [mag] & &}
\startdata
16295928-2410106 & 247.497037 & -24.169739 &  M4.75 & 10.03 & 0.23 & debris &  L1709\\
16302673-2359087 & 247.611402 & -23.985908 & M6     & 11.48  & 0.12 & full & L1709 \\
16303059-2413373 & 247.627503 & -24.227148 & M8     &12.71   & 0.13 & nodisk & L1709\\
16310240-2408431 & 247.759966 & -24.145469 & M5     &10.79   &  0.14 & full & L1709\\
16310516-2404401 & 247.771483 & -24.077977 & M0     & 8.58   & 0.29 & full & L1709\\
16310650-2403000 & 247.777038 & -24.050159 & M5     & 9.37   & 0.22 & debris & L1709\\
16321089-2359080 & 248.045428 & -23.985708 & M2     & 10.15  & 0.80 & nodisk &L1709 \\
16261886-2428196 & 246.578617 & -24.472256 & M0     & 8.07   & 1.83 & full & L1688 \\
16272183-2443356$^{b}$ & 246.840921 & -24.726633 & M3     & 10.78  & 4.00 & nodisk & L1688\\
16265429-2424381 & 246.726254 & -24.410589 & M4     & 13.79  & 6.07 & evolved & L1688\\
16261898-2424142 & 246.579096 & -24.403978 & M6     & 11.94  & 2.86 & full&  L1688\\
16265733-2435388 & 246.738871 & -24.594133 & M6     & 12.81  & 3.80 & evolved & L1688\\
16265197-2430394 & 246.716525 & -24.511006 & M5     & 13.46  & 5.09 & full & L1688\\
16262523-2423239$^{b}$ & 246.605133 & -24.390006 & M5.5   & 13.33  & 6.62 & nodisk & L1688\\
16273863-2438391 & 246.910918 & -24.644337 & M7.5  & 11.08 & 0.70 &  full & L1688 \\
16314581-2439089 & 247.940892 & -24.652616 & M7.75 & 11.03  & 0.61 & full & L1689\\
\enddata
$^{a}$ Full disk: optically thick with no large holes; evolved disk: optically thin with no large hole, and debris disk: second generational dust from planetesimal collisions.\\
$^{b}$ Potential candidates to present spherical mass loss based on overluminosity and disk type.
\end{deluxetable*}

\end{document}